\documentclass[journal]{IEEEtran}

\usepackage[colorlinks,bookmarksopen,bookmarksnumbered,citecolor=blue,linkcolor=blue,urlcolor=blue]{hyperref}
\usepackage{graphicx}
\usepackage{amsmath}
\usepackage{array}
\usepackage{booktabs}
\usepackage{multirow}
\usepackage{graphicx}%
\usepackage{subcaption}

\ifCLASSINFOpdf
\else
\fi
\usepackage{enumerate}

\begin{document}
%
\title{
 M3S-UPD: Efficient Multi-Stage Self-Supervised Learning for Fine-Grained Encrypted Traffic Classification with Unknown Pattern Discovery

}
%
%
%

\author{Yali~Yuan,
        Yu~Huang,
        Xingjian~Zeng,
        Hantao~Mei,
        and~Guang~Cheng*,~\IEEEmembership{Member,~IEEE}
\thanks{Yali Yuan, Yu Huang, Xingjian Zeng, Hantao~Mei, and Guang Cheng (the corresponding author) are with the School of Cyber Science and Engineering, Southeast University, Nanjing 211189, China. (Corresponding author’s e-mail: chengguang@seu.edu.cn).}
}

%
%

\markboth{Journal of \LaTeX\ Class Files,~Vol.~14, No.~8, August~2015}%
{Shell \MakeLowercase{\textit{et al.}}: Bare Demo of IEEEtran.cls for IEEE Journals}
%



\maketitle

\begin{abstract}
The growing complexity of encrypted network traffic presents dual challenges for modern network management: accurate multiclass classification of known applications and reliable detection of unknown traffic patterns. Although deep learning models show promise in controlled environments, their real-world deployment is hindered by data scarcity, concept drift, and operational constraints. This paper proposes M3S-UPD, a novel Multi-Stage Self-Supervised Unknown-aware Packet Detection framework that synergistically integrates semi-supervised learning with representation analysis. Our approach eliminates artificial segregation between classification and detection tasks through a four-phase iterative process: 1) probabilistic embedding generation, 2) clustering-based structure discovery, 3) distribution-aligned outlier identification, and 4) confidence-aware model updating. Key innovations include a self-supervised unknown detection mechanism that requires neither synthetic samples nor prior knowledge, and a continuous learning architecture that is resistant to performance degradation. Experimental results show that M3S-UPD not only outperforms existing methods on the few-shot encrypted traffic classification task, but also simultaneously achieves competitive performance on the zero-shot unknown traffic discovery task.
\end{abstract}


\IEEEpeerreviewmaketitle

\section{Introduction}
Nowadays, a vast number of network applications that employ encrypted traffic for communication continuously emerge, leading to an increasingly complicated and diverse network environment.
The expanding types of traffic, coupled with the deployment of encryption methods for privacy preserving, pose challenges to network management and censorship.
This not only further underscores the critical role of encrypted traffic classification, but also elevates practical demands on its applications.

From an application-oriented perspective, encrypted traffic classification (ETC) can be divided into two sub-tasks: 
\begin{itemize}
    \item The multi-classification task aimed at identifying various traffic types. For example, controlling the quality of service (QoS) and allocating network resources require accurate and robust traffic classification.
    \item The detection task focused on discovering unknown traffic that has not been observed by the classifier. In practical network management scenarios, intrusion detection and malicious traffic classification can be formulated as unknown traffic detection tasks.
\end{itemize}
Existing Machine Learning (ML) and Deep Learning (DL)-based ETC methods have proven their effectiveness on the two tasks in laboratory settings. 
These methods focus on efficient, accurate, and robust encrypted traffic classification by leveraging the powerful feature extraction and learning capabilities of DL-based classifiers.
Although these methods demonstrate notable classification performance, some practical issues remain to be discussed.
\begin{enumerate}[1)]
    \item Data Scarcity.
    Obtaining abundant labeled encrypted traffic is challenging in the online learning context.
    The insufficiency of training data can lead to limited performance of classifier.
    However, most DL-based models designed for ETC are trained in a supervised manner, with an unrealistic assumption that substantial labeled training data can be obtained at a low cost (in terms of time and manpower). 
    Such scarcity in training dataset popes great difficulties to the actual application of DL-based models for online encrypted traffic classification. 
    \item Concept Drifting. 
    Learning classifiers from real-world traffic encounters the change in distribution and characteristics of traffic, whose hidden data contexts and labels may vary and become unknown to the model. 
    Such phenomenon is known as concept drifting. In online traffic classification, the fluctuation of traffic labels is a common and challenging type of concept drifting.
    The varying labels of online traffic not only demands the classifier to be updatable, but also efficiency in model training and deploying, which is challenging for DL-based models with significant training costs.
    Such issues become increasingly pressing with the continuous growth in the types of network traffic. 
    
    \item Model Limitations. 
    Existing ETC techniques achieve commendable performance in specific laboratory settings, yet still exhibit limitations in a real-world scenario.
    To address challenges such as data scarcity and concept drifting, data augmentation techniques and self-paced learning are employed to generate pseudo-labels for unlabeled data and simulate the distribution of unknown traffic.
    However, these methods face multiple practical challenges.
    GAN-based methods leverage the easily-obtained unlabeled traffic data to boost classification performance, but their heavily relies on the selection of hyper-parameters may limit the general application in real-time and ever-changing online network environments.
    Models that leverage self-supervised mechanism enlarge the labeled training set and detect unknown traffic based on model-learning results and sample verification, but potentially suffer from low confidence of pseudo labels, unreliable verification of samples, and potential update inaccuracies. 
\end{enumerate}

Overall, multi-class traffic classification and unknown traffic detection are comprehensive tasks faced with numerous challenges stemming from model limitations and intrinsic properties of real-world traffic.
These challenges require models to possess strong learning capabilities, efficient updates, and convenient deployment.
Furthermore, though always studied separately by existing approaches, the multi-classification task and detection task of online encrypted traffic are strongly interconnected for the following reasons.
First of all, they are both classification problems, only with different classifying outcomes.
Second, the superior performance of classifier for both tasks relies on the comprehensive understanding of training data and the good transferability on acquired knowledge, which indicates the similar inherent properties of the two different application-oriented tasks.
Finally, in an open-world online context, traffic continuously flowing towards the classifier inevitably contains unseen categories, which indicates a reasonable and feasible application scenario for detecting unknown traffic while classifying the known ones.
The inherent correlations of encrypted traffic classification and unknown traffic discovery imply a possible unified approach for addressing these two tasks simultaneously.

Based on the above considerations, we propose a self-supervised training framework for encrypted traffic classification and unknown traffic detection under the condition of limited labeled training data in this paper.
Starting with a suboptimal classification model trained on limited labeled data, which cannot identify unknown traffic, the proposed training framework aims at incrementally boosting the original model's performance and gradually achieving accurate unknown traffic detection via reasonable utilization of unlabeled traffic data.

Instead of using data augmentation methods to synthesize unknown traffic samples for training the model, the proposed framework does not rely on any prior knowledge of unknown traffic and achieves a unified classification of known/unknown traffic through multiple training steps, each consisting of four stages.
In the \textbf{model preparing} stage, a recently updated classification model is used for generating classification probability distribution and data embeddings for unlabeled traffic.
Subsequently, in the \textbf{embedding clustering} stage, data embeddings of unlabeled traffic is clustered and divided to distinct categories, representing the spatial distribution of unlabeled traffic in the embedding space.
Later, the unlabeled traffic data in different clusters are aligned with known classes of the training set in the \textbf{spatial distribution aligning} stage, and assigned with corresponding auxiliary labels.
Samples fail to align will be initially classified as potential unknown traffic.
Finally, a consistency-check between classification probabilities and aligning outcomes of unlabeled traffic is conducted for \textbf{reliable model updates}.
The training dataset is expanded with unlabeled samples with highly confident pseudo-labels, while unlabeled traffic samples that fail to align and have abnormally low predicted classification probabilities are identified as unknown traffic.
The model is then updated on the expanded dataset and proceeds to the next training step.

To conclude, this paper mainly contributes in three aspects:
\begin{itemize}
    \item We propose a novel self-supervised training framework, M3S-UPD, for encrypted traffic classification in a limited labeled training data scenario, which poses a challenge for existing DL-based methods. By gradually expanding labeled training data with highly confident pseudo-labels of unlabeled traffic, the performance of the initially suboptimal model is incrementally improved.

    \item We enable the classification model for known traffic categories to detect unknown traffic classes without any prior knowledge and data augmentation.
    Through consistency analysis of embedding-level spatial distribution and model-level predicted outcomes, samples with discrepancies between clustering patterns and classification probabilities are accurately and efficiently identified as unknown traffic.
    \item 
    We conduct comprehensive experiments to evaluate our proposed method on two public experimental datasets. 
    The proposed M3S-UPD demonstrates competitive performance compared to state-of-the-art methods with limited training data in both closed world and open world settings.
    Furthermore, extensive experiments incorporating moderate expert knowledge show that M3S-UPD achieves fine-grained traffic classification and effectively adapts to frequently updated datasets, where unknown traffic classes are continuously identified, labeled, and added.
\end{itemize}

The remainder of the paper is structured as follows: 
Section~\ref{relatedwork} reviews related literature. 
Section~\ref{Problem Definition} defines the problem and outlines three key challenges.
Section~\ref{Method} introduces our self-supervised learning framework for online encrypted traffic classification and unknown traffic discovery. 
Section~\ref{section5} details the results of our evaluation.
Section~\ref{Conclusion} offers the conclusion of the study.

\section{Related Work}
\label{relatedwork}

\subsection{Traffic Classification on Encrypted Traffic}

Traffic Classification (TC) pertains to the task of associating user traffic with the applications, services, and software generating them. This is widely employed for various purposes, including quality-of-service (QoS) \cite{Barradas2021FlowLensEE,Yun2023EncryptedTT} , network management \cite{Wang2023ATA,Li2022PacketLevelOA}, and intrusion detection and defenses \cite{Fu2023FrequencyDF}. Over the past decades, numerous TC methods have been proposed, such as port-based and deep packet inspection (DPI) methods \cite{Bai2008TrafficIO}  that utilize default port numbers and application signatures. However, these methods have become less effective due to the proliferation of network address translation (NAT) \cite{Sicker2007LegalIS}  and packet encryption \cite{Finsterbusch2014ASO}.

Many machine learning methods have been introduced to build traffic classifiers by extracting implicit patterns. AlSabah et al. \cite{AlSabahBG12}  extracted features like circuit lifetime, data transferred, cell inter-arrival times, and the number of cells sent recently. They utilized Naïve Bayes, Bayesian Networks, and Decision Trees to classify browser, P2P, and media traffic. Cuzzocrea et al. \cite{Cuzzocrea2017TorTA}  employed Mann-Whitney test and Kolmogorov-Smirnov test to verify the significant difference between the distribution of Tor traffic and normal traffic features. Subsequently, they used machine learning algorithms to classify seven Tor traffic types. Montieri et al. \cite{Montieri_Anonymity}  extended Tor traffic classification to the application level. Xu et al. \cite{PathSignature22}  transformed packet sequences into paths for the classification of multiple encrypted traffic data. Some research utilized KNN and SVM algorithms to identify websites of traffic, known as Website Fingerprinting \cite{WFatIS16,Cherubin2022OnlineWF} .

In recent years, scholars have introduced deep learning for encrypted traffic classification. Liu et al. \cite{Liu_FSNet}  input the packet length sequence into a model using encoders, decoders, and RNN to extract features and achieve TLS traffic classification. Wang et al. \cite{Wang2023ATA}  conducted fast traffic classification with a Temporal Convolutional Network (TCN). Specifically, fast classification was performed for flows accurately classified with only the first few packets, while complex flows were analyzed in detail. Their method enables efficient encrypted traffic classification by extracting the payload length of packets and constructing a TCN classifier. Zhao et al. \cite{Zhao2022FlowSA}  considered the flow sequence as a graph, constructing the graph structure with feature vectors including application and time. They then extracted features using four residual graph neural network (ResGCN) modules and a 3-layer multilayer perceptron to achieve traffic classification. Several research building on deep learning, introduced new mechanisms such as the self-attention mechanism, multi-level self-attention, transformer, and ensemble learning mechanism to further improve classification performance \cite{Yun2023EncryptedTT,Zhao2023YetAT,Deng2023RobustMW,snWF22} . Although these methods have shown high performance on experimental datasets, it's important to note that these models are fixed and cannot recognize classes that have not been learned.

\subsection{Model Update of Traffic Classifiers}

To tackle the challenge of model inflexibility, several studies aim to enhance the model's ability to generalize, enabling it to recognize samples it hasn't encountered during training. Ede et. al., \cite{Ede2020FlowPrintSM} . devised a semi-supervised encrypted traffic classification system. They clustered traffic of different application types into distinct clusters based on time, device, and destination features, constructing an app fingerprint for traffic classification. Being unsupervised, their method can identify apps not explicitly trained. Fu et al\cite{Fu2023DetectingUE} . similarly employed unsupervised learning for identifying unknown traffic. They transformed interaction patterns between long and short flows into graph structural features, detecting encrypted malicious traffic by analyzing graph connectivity and sparsity.
Lifelong machine learning empowers models to continuously accumulate new knowledge, saving classifier training costs and mitigating concept drift. Attarian et al\cite{Attarian2019AdaWFPAAO} . proposed AdaWFPA, an adaptive online website traffic recognition method. Upon the arrival of new training samples, the model predicts, compares with true labels, and updates based on the results. Zhang et al\cite{Zhang2020AutonomousUF} . introduced a self-updating model framework that judges unknown class packets based on the classifier's results, compares them with existing knowledge, and annotates to form a new dataset for updates. The authors further discuss trigger times for model updates in \cite{Fu2023DetectingUE} , based on classification output instability, enhancing the update framework's performance. Current model updates often rely on inferring unknown traffic from existing knowledge, facing challenges of initial dataset comprehensiveness and potential update inaccuracies.

\subsection{Few-Shot Traffic Classification}

Due to the challenge of obtaining abundant encrypted training data, several studies are dedicated to addressing the sparsity issue in training datasets. Wang et al\cite{Wang2016OnRA} . assessed the minimum training set needed to achieve high-performance website fingerprinting. Remarkably, their research demonstrated that a mere 6,800 samples could maintain highly accurate recognition for 100 websites. Sirinam et al\cite{Payap2019TFMP} . introduced Triplet, a method that initially learns traffic distribution knowledge from a substantial amount of non-target traffic. Subsequently, it utilizes a small number of target traffic samples to train the target classifier, achieving website identification with only 5 samples. Oh et al\cite{Oh2021GANDaLFGF} . leveraged generative adversarial networks to generate a substantial amount of "fake" data from a limited set of training samples, aiding in training deep neural network classifiers. Zhou et al\cite{Zhou2023FewshotWF} . proposed a website fingerprinting attack method capable of updating the classifier with a small number of new samples. This approach maps samples to the deep learning feature space, clusters data samples based on training labels, and aligns the clustering center of new samples with that of training samples, facilitating classifier updates. Hu et al\cite{Hu2022AttributeBasedZL} . designed an attribute-based zero-shot encrypted traffic classification framework. They used a Temporal Convolutional Network (TCN)-based feature embedding model and a Simple Recurrent Unit (SRU)-based attribute embedding model to transform traffic into joint embeddings of attribute values. The framework employs a Generative Adversarial Network (GAN)-based feature generation model for recognizing unknown classes. While small sample learning significantly reduces training costs, some challenges persist, including reliance on knowledge from the original training set for judgments and learning from newly arrived samples, leading to verification difficulties and potential update inaccuracies.
\section{Problem Definition}
\label{Problem Definition}

This section defines the problem scenario of traffic multi-class classification and unknown traffic identification when only limited training data is available. Let $D=\{(x_1,y_1),(x_2,y_2),\ldots,(x_n, y_n)\} $ be the training dataset that contains $M$ known traffic classes, i.e., $y_i \in C=\{ l_1,l_2,..,l_M \}$ where $x_i$ represents the input feature vector corresponding to training sample in traffic class $y_i$ and  $C$ is the set of labels for known traffic categories.
The classifier trained on dataset $D$ with scarce data  potentially suffers from limited performance on mapping input vector $x_i$ to the traffic label $y_i$ while lacking the ability to identify unknown traffic patterns.

At time $t$, suppose the classifier $F^t$ has completed its training on the previous dataset $D^{t-1}$ with known label set $C^{t-1}$. 
For the incoming traffic flows $N^t = K^t \cup U^t$, $K^t$ represents the set of traffic with known labels and $U^t$ represents the set of traffic whose labels haven't appeared in the known label set $C^{t-1}$. 
The classifier's task is to predict the specific classes of the traffic samples from $K^t$, i.e., mapping the input vectors from $K^t$ to certain labels in $C^{t-1}$ and to find any traffic categories that have not appeared in $D^{t-1}$ from $N^t$, i.e., distinguish the traffic in $U^t$ from those in $K^t$.

To move on, the classifier needs to be updated in time for classifying traffic and identifying unknown categories in the coming traffic flows at $t+1$. 
To be specific, a new training dataset $D^t=\{D^{t-1} \cup {K^{t}}^{'}|{K^{t}}^{'} \in K^t$ \} is required for updating the classifier $F^t$.
If possible, incorporating expert knowledge can further enlarge training dataset by labeling the previously unknown traffic categories and merge them to the known label set $C^t$. 
In this way, the training dataset is expanded as $D^t=\{D^{t-1} \cup {K^{t}}^{'} \cup {U^t}^{'}|{K^{t}}^{'} \in K^t,  {U^{t}}^{'} \in U^t\}$.

At this point, three subproblems emerge as we wish for the classifier to continuously handle online traffic in a sustained, robust and effective manner:

\begin{enumerate}
    \item Limited training data. 
    Obtaining substantial training data can be unrealistic due to the challenges of data labeling
    When only limited training data is available, classifiers may struggle to effectively manage the continuous influx of online traffic. 
    Therefore, it is crucial to develop strategies for gradually expanding $D^t$ enhance model performance.
    \item Detection of unknown traffic.
    The classifier $F^t$ is trained on dataset $D^{t-1}$ with a known traffic label set $C^{t-1}$, making the transfer of this acquired knowledge to unknown traffic a considerable challenges.
    It is essential to devise methods for recognizing new traffic patterns by leveraging insights gained from the locally labeled training data.
    \item Reliable model updates. 
    Enabling the model to achieve incremental known traffic classification as well as effective unknown traffic detection performance necessitates reliable model updates throughout the training process.
    On one hand, ensuring high confidence in newly assigned labels when utilizing unlabeled traffic to expand the labeled dataset poses challenges due to potential biases from the model's self-learning.
    On the other hand, the lack of prior knowledge about unknown traffic complicates the effective transfer of knowledge acquired from known traffic data to achieve accurate and low false-positive unknown traffic detection.
    Additionally, employing data augmentation techniques such as GANs to simulate unknown traffic training data results in high training costs, potentially inaccurate data estimates, and unstable model performance.
\end{enumerate}

\section{Method}
\label{Method}

\begin{figure*}[!htb]
    \centering
    \includegraphics[width=1\linewidth]{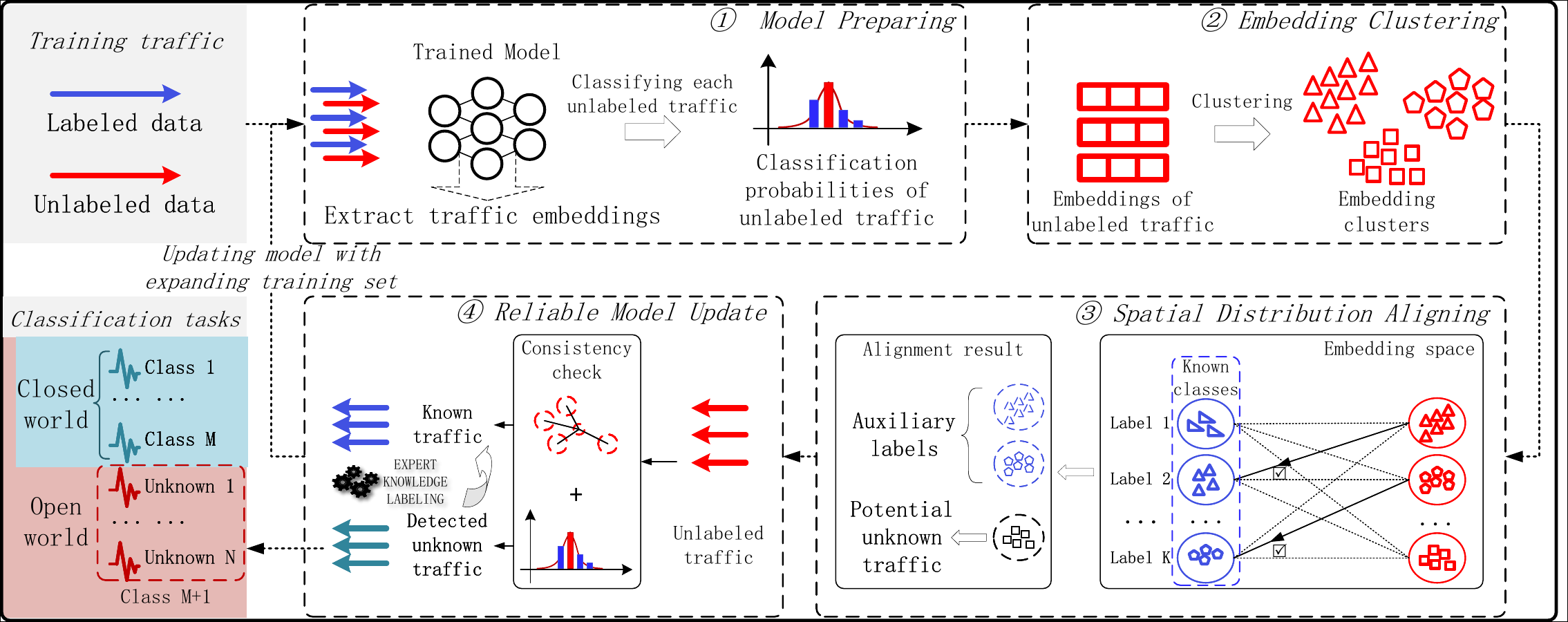}
    \caption{Overview of the proposed method.}
    \label{fig:system}
\end{figure*}
This section provides a detailed introduction to the proposed framework (as shown in Fig. \ref{fig:system}) designed to address these three subproblems.
With only scarce labeled data to learn from, a classification model with constrained performance is trained, temporarily failing to identify unknown traffic.
Leveraging the self-supervised learning paradigm, we aim to expand the labeled training data with high confidence for model updates, enabling accurate known traffic classification with facilitating unknown traffic detection after a fixed number of training steps.
Accordingly, a training framework consisting of four stages is presented.

In the first stage, the model trained in the previous step takes unlabeled traffic as input to generate classification probabilities and data embeddings.
Subsequently, these embeddings are used to generate embedding-space clusters that reflect the spatial distribution of unlabeled traffic.
By calculating distances between these embedding clusters and each known class in the training dataset, unlabeled traffic is temporarily aligned with a specific known class, receiving corresponding auxiliary labels.
Finally, a consistency check is performed between the classification probabilities distribution and the cluster-level auxiliary labels to ensure reliable model updates. 
The labeled training data is supplemented with samples whose auxiliary labels match with the highest classification probabilities. 
Meanwhile, unlabeled traffic that fails to align and exhibits low classification probabilities is identified as unknown traffic.
If necessary, these detected unknown traffic can be labeled with the assistance of expert knowledge and added to the training set for further fine-grained traffic classification.

\subsection{Model Preparing}
As we consider a realistic scenario where only limited labeled traffic data is available, existing DL-based traffic classification methods with intrinsic data-intensive training processes become inapplicable.
Therefore, based on the self-supervised learning paradigm, we aim to conduct high-confidence expansion of labeled training data for model updates to realize accurate known traffic classification with further unknown traffic discovery by exploiting sample classification probability distribution and data embedding characteristics.

In the training pipeline guided by such idea, only a small amount of labeled data is available at the beginning.
We aim to initially train a weak model then gradually boost its performance on known traffic multi-classification and unknown traffic detection.
During the training process, the model receives data samples with labels as inputs and maps raw features to known classes' classification distribution by minimizing standard cross-entropy loss:
\begin{equation}
    \mathcal{L}_s(X)=\frac{1}{N} \sum_{i=1}^{N}H(y_i, y^{'}_{i}),
    \label{eq: ce}
\end{equation}
where $y_i$ and $y_{i}^{'}$ respectively refers to the ground-truth label and predicted label of input data sample.

Once finishing loss minimization on labeled training data, the model can generate classification probabilities and data embeddings of given input traffic.
Data embeddings are vectors generated by transforming the original raw data inputs into lower-dimensional representations for the model.
With optimized model network parameters, the input traffic data can be transformed into lower-dimensional embeddings that well capture the inner characteristics of traffic sequences.
By applying linear transformation on these embeddings, a raw classification score vector of each class $\mathbf{z}=[z_1, z_2, \ldots, z_K]$ is produced and later normalized with a softmax function to generate final classification probabilities distribution:
\begin{equation}
\sigma(\mathbf{z})_i = \frac{e^{z_i}}{\sum_{j=1}^{K} e^{z_j}} \quad \text{for } i = 1, \ldots, K
    \label{eq: softmax}
\end{equation}
To some extent, the maximum classification probability of input traffic sample reflects the confidence that it belongs to according known class.

\subsection{Embedding Clustering}

To tackle with limited training data, clustering technique is utilized for aiding traffic labeling from an embedding perspective.
After generating data embeddings of unlabeled traffic with trained model, 
a clustering process is conducted to divide unlabeled traffic into different cluster categories in the embedding space, instead of the original input space.
Data embeddings are high-level features generated by a trained model with a specific architecture and optimized parameters, which are utilized to derive the final predicted probability distribution.
Therefore, the spatial distribution of embeddings obtained by the clustering process is highly related to the predicted probability distribution of the model, and the two can be integrated to expand labeled training data while detecting unknown traffic with high confidence.

Specifically, unlike previous researches that apply K-Means clustering algorithm, unlabeled traffic data is divided into different clusters with regard to density distribution by DBSCAN.
K-Means algorithm relies on manually selected hyperparameter for specifying the number of clusters, which is difficult to determine based on prior knowledge in the process of real-world traffic classification.
In real-world traffic classification scenario, expert knowledge can be cooperated to identify and label newly discovered traffic for updating local dataset, resulting in dynamically varying sample labels of training data.
Considering fine-grained traffic classification and continuous model updates, clustering unlabeled traffic into a predefined number of clusters may result in suboptimal clustering outcomes that inaccurately describes the spatial distribution of data.
Henceforth, DBSCAN is chosen for clustering traffic data embeddings due to its property of automatically determining the number of clusters.

\subsection{Spatial Distribution Aligning}
To reliably label unlabeled data in a self-supervised manner for expanding training dataset, traffic data embeddings are temporarily mapped to a certain known traffic class in the training set by an aligning mechanism with corresponding auxiliary labels.
For a unlabeled traffic embedding cluster $u_i$ and embeddings of a class in the training dataset $k_m$, the distance between these two data embedding segments is computed as:
\begin{equation}
    d(u_i, k_m)= ||v_{i} - \mu_{m}|| ^ 2,
    \label{eq: distance}
\end{equation}
where $v_{i}$ and $\mu_{m}$ represents the centroids of unlabeled embedding cluster $u_i$ and labeled data embedding of class $m$  respectively.

By calculating such pair-wise distances between each cluster in unlabeled traffic embeddings and each class in labeled traffic embeddings, the auxiliary labels $\widetilde{y}_i$ of unlabeled traffic data in cluster $u_i$ is determined as follows.
\begin{equation}
    \widetilde{y}_i=\left\{
    \begin{aligned}
    & Potential \ unknown & \quad if \ \mathop{\arg\min}_{m} \ d(u_i, k_m) \geq t \\
    & \mathop{\arg\min}_{m} \ d(u_i, k_m) &  \quad otherwise ,
    \end{aligned}
    \right.
    \label{eq: SDAligning}
\end{equation}
where $t$ is a distance threshold. 
Any clusters of unlabeled traffic data embeddings with a minimum distance from all known classes exceeding a certain threshold $t$ is considered potential unknown traffic and will be excluded from alignment with any known class labels in the training dataset.
Otherwise, samples within the clusters of unlabeled traffic data embeddings will be temporarily assigned the label of the known class that is closest to them.




\subsection{Reliable Model Update}
In the final stage of the training framework, the current model undergoes reliable updates, which consist of incremental improvements in two performance aspects:
\begin{itemize}
    \item Classification performance on known traffic. 
    Due to the limited amount of data used for training, the initial model exhibits constrained performance in classifying known traffic. To address this, the unlabeled traffic in the training data is reliably labeled based on the model's learning outcomes and the distribution patterns of traffic data embeddings. These newly labeled samples are then added to the training set to update the model. During the subsequent training process, the classification performance of the model will improve incrementally as the labeled dataset continues to expand.
    \item Detection performance on unknown traffic. 
    Unlike previous methods that leverage complex and challenging-to-train adversarial generative networks to construct unknown traffic samples for model training, the detection of unknown traffic achieved in the proposed framework is an efficient and low-overhead consequence of the model's reliable updates. Rather than enabling the model to classify unknown traffic by learning from synthetic samples, the detection capability of the proposed framework arises from the consistency check between the model's learning outcomes and the distribution of unlabeled traffic data embeddings.
\end{itemize}
For a given unlabeled traffic data sample, the model outputs the predicted classification probabilities for each known class. The highest prediction probability can be considered an estimate of the model's prediction confidence. In self-supervised learning methods based on self-labeling, assigning labels corresponding to the highest predicted probabilities to unlabeled samples is a common approach for expanding the training set.
However, such approach potentially suffer from biases in the model's self-learning process, leading to newly labeled samples with low confidence, which ultimately results in inaccurate model updates and suboptimal performance.
Therefore, a consistency check between the confidence of unlabeled traffic data and its corresponding aligning outcomes is introduced for guaranteeing reliable self-labeling.

For successfully aligned unlabeled data, a sample will only be added to the training set under the corresponding known traffic class if its auxiliary label matches the label assigned by the model to its highest predicted classification probability.
While focusing on high-confidence unlabeled samples to achieve growth in the labeled dataset, efficient identification of unknown traffic is achieved through the consistency check between low-confidence samples and their alignment results.
For those samples that fail to align and are considered potential unknown traffic, if their confidence, represented by the model's highest predicted classification probability, falls within the lowest range of the overall unlabeled data distribution, these samples are identified as belonging to the unknown traffic class.

To improve the efficiency and effectiveness of the consistency check, the unlabeled data is sorted by the model's classification confidence, and the top $t_{top}$ and bottom $t_{bottom}$ samples, with the highest and lowest confidence, respectively, are selected as candidates for consistency checks. Other unlabeled samples, with confidence scores falling in the middle of the distribution, are not processed further as they do not exhibit a clear known/unknown distinction in the model's predictions, and are deferred for checking in the next training iteration.

The underlying motivation for such consistency checks stems from the idea that the model’s prediction confidence to some extent reflects whether a traffic data sample belongs to a known class or not. 
This is because the model is never trained with unknown traffic during updates, leading to a lack of strong classification ability for unknown traffic. 
As a result, the lower the prediction confidence for an unlabeled sample, the higher the likelihood that it belongs to an unknown class, while higher prediction confidence suggests that the sample is likely associated with a known class.
By incorporating knowledge of the data embedding distribution, bias that may arise from this self-learning judgment is greatly mitigated by the alignment results of the unlabeled traffic data.
Furthermore, taking into account real-world unknown traffic detection scenarios, the proposed framework allows for the introduction of expert knowledge to label unknown traffic identified during the training process. This enables the expansion of the local training set with new traffic categories, facilitating continuous fine-grained model updates to adapt to complex network environments and the ongoing emergence of new traffic types.

\section{Experimental Evaluation}
\label{section5}

\subsection{Dataset and Experimental Setup}
In this section, we conduct extensive experiments to demonstrate the efficacy of our proposed M3S-UPD in traffic classification. Our experiments utilize the widely recognized Tor public dataset, ISCXTor2016~\cite{lashkari2017characterization}, curated by Lashkari and a Tor dataset that we collected ourselves, TDTor. The ISCXTor dataset comprises over 8000 Tor samples spanning 8 traffic types: VoIP, P2P, FILE-Transfer, Browsing, Video, Mail, Audio, and Chat amounting to a total size of 22.8 GB with 85 PCAP files. The Tor dataset we collected comprises over 12000 Tor samples spanning 7 traffic types. We removed File-Transfer traffic because this type of traffic is not common in Tor~\cite{chaabane2010digging}. The distributions of ISCXTor and TDTor are presented in Table~\ref{table: Category Distribution in ISCXTor Dataset} and Table~\ref{table: Category Distribution in TDTor Dataset}, respectively. It can be observed that TDTor exhibits a more balanced distribution compared to ISCXTor. The evaluation is conducted under a realistic scenario where the attacker has only partial samples from certain categories for model training but needs to identify samples from all categories, including unknown traffic. The original datasets ISCXTor and TDTor were partitioned into training, validation, and test sets with a ratio of 6:2:2. To investigate the effectiveness of our proposed method in various scenarios of known/unknown attacks detection, we constructed two scenarios with two data settings for both datasets as shown in Table~\ref{table:setting}.
\begin{itemize}
    \item No-Expert: This scenario evaluates the recognition ability of the model without introducing expert knowledge. In this scenario, all non-known classes are categorized into unknown classes. In this scenario, initially only 30\% of each known class is selected for model training.
    \item With-Expert: This scenario evaluates the complete recognition capability of our proposed model, which continuously learns knowledge from new classes by introducing expert knowledge. In this scenario, initially only 30\% of each known class is selected for model training.
    \item Setting1: We select three types as known classes and other types as unknown classes. For the known classes, we select 30\% of the total samples as training samples. This scenario explores the model's recognition ability when only a few known classes are available.
    \item Setting2: We select five types as known classes and other types as unknown classes. For the known classes, we select 30\% of the total samples as training samples. This scenario explores the model's recognition ability when most known classes are available and only a few unknown traffic classes are unavailable.

\end{itemize}

\begin{table}[h!]
\centering
\begin{tabular}{lcc}
\toprule
\textbf{Category} & \textbf{Count} & \textbf{Percentage} \\
\midrule
Audio & 1026 & 7.1\% \\
Browsing & 2645 & 18.2\% \\
Chat & 485 & 3.3\% \\
FILE-Transfer & 1663 & 11.5\% \\
Mail & 497 & 3.4\% \\
P2P & 2139 & 14.7\% \\
Video & 1529 & 10.5\% \\
VOIP & 4524 & 31.2\% \\
\midrule
\textbf{Total} & 12808 & 100\% \\
\bottomrule
\end{tabular}
\caption{Category Distribution in ISCXTor Dataset}
\label{table: Category Distribution in ISCXTor Dataset}
\end{table}

\begin{table}[h!]
\centering
\begin{tabular}{lcc}
\toprule
\textbf{Category} & \textbf{Count} & \textbf{Percentage} \\
\midrule
Audio & 1474 & 5.4\% \\
Broswer & 5000 & 18.3\% \\
Mail & 5000 & 18.3\% \\
Message & 3597 & 13.1\% \\
P2P & 5000 & 18.3\% \\
Video & 2314 & 8.4\% \\
VOIP & 5000 & 18.3\% \\
\midrule
\textbf{Total} & 27385 & 100\% \\
\bottomrule
\end{tabular}
\caption{Category Distribution in TDTor Dataset}
\label{table: Category Distribution in TDTor Dataset}
\end{table}

\begin{table*}[!htb]
\caption{Experimental data settings.}
\label{table:setting}
\centering
\begin{tabular}{|c|cc|cc|cc|cc|}
\hline
\multirow{2}{*}{}        & \multicolumn{2}{c|}{Setting1 no Expert}       & \multicolumn{2}{c|}{Setting 1 no Expert}      & \multicolumn{2}{c|}{Setting2 with Expert}     & \multicolumn{2}{c|}{Setting 2 with Expert}    \\ \cline{2-9} 
\multirow{5}{*}{ISCXTor} & \multicolumn{1}{c|}{Known}         & Unknown  & \multicolumn{1}{c|}{Known}         & Unknown  & \multicolumn{1}{c|}{Known}         & Unknown  & \multicolumn{1}{c|}{Known}         & Unknown  \\ \hline
                         & \multicolumn{1}{c|}{VOIP}          & Browsing & \multicolumn{1}{c|}{VOIP}          & Browsing & \multicolumn{1}{c|}{VOIP, Video}          & Browsing & \multicolumn{1}{c|}{VOIP, Video}          & Browsing \\
                         & \multicolumn{1}{c|}{P2P}           & Video, Mail    & \multicolumn{1}{c|}{P2P}           & Video, Mail     & \multicolumn{1}{c|}{P2P, Chat}           & Mail    & \multicolumn{1}{c|}{P2P, Chat}           & Mail     \\
                         & \multicolumn{1}{c|}{FILE-Transfer} & Audio, Chat     & \multicolumn{1}{c|}{FILE-Transfer} & Audio, Chat    & \multicolumn{1}{c|}{FILE-Transfer} & Audio     & \multicolumn{1}{c|}{FILE-Transfer} & Audio    \\ \hline
\multirow{3}{*}{TDTor}   & \multicolumn{1}{c|}{Browser}      & VoIP     & \multicolumn{1}{c|}{Browser}       & VoIP     & \multicolumn{1}{c|}{Browser}      & Video     & \multicolumn{1}{c|}{Browser}       & Video     \\
                         & \multicolumn{1}{c|}{Mail}          & Message  & \multicolumn{1}{c|}{Mail}          & Message  & \multicolumn{1}{c|}{Mail, P2P}          & Audio  & \multicolumn{1}{c|}{Mail, P2P}          & Aduio  \\
                         & \multicolumn{1}{c|}{P2P}           & Video, Audio    & \multicolumn{1}{c|}{P2P}           & Video, Audio    & \multicolumn{1}{c|}{VoIP, Message}           &     & \multicolumn{1}{c|}{VoIP, Message}           &     \\ \hline
\end{tabular}
\end{table*}

Consequently, we have two scenarios named no Expert and with Expert, and four new datasets named by ISCXTor setting1, ISCXTor setting2, TDTor setting1 and TDTor setting2. 
We selected several state-of-the-art (SOTA) methods as evaluation baselines due to their strong performance in previous work. which are referred as CVAE-EVT~\cite{Yang2021ConditionalVA}, Cls-Anomaly~\cite{Yang2021ConditionalVA} and EVM~\cite{baseline2henrydoss2017incremental}. 
\begin{itemize}
    \item CVAE-EVT proposes an intelligent intrusion detection method which can classifying known attacks as well as inferring unknown ones. It enables high-performance hierarchical attacks detection by minimizing the empirical risk and open-set risk.
    \item Cls-Anomaly proposes an anomaly detection model which assemble a classification model and an anomaly detection model. A random forest-based classification model classifying a flow as benign or one of known attacks and a SVM based anomaly detection model identifying whether this classification is correct or not. 
    \item EVM, as known as Extreme Value Machine, is novel open-set designed classifier that supports variable bandwidth incremental learning. This method utilize the EVM for intrusion detection and measure the open set recognition performance of identifying known and unknown classes.

\end{itemize}

We first demonstrated the classification performance of the above four data settings on the ISCXTor and TDTor dataset in Section~\ref{subsec: perfor-eval} to evaluate the effectiveness of our proposed method for traffic classification, particularly its ability to handle unknown traffic. Subsequently, in Section~\ref{subsec:compare-sota}, we compare our method with multiple SOTA methods to evaluate its effectiveness. Finally, we conducted ablation experiments in Sections~\ref{sub:diff-num-known} and~\ref{subsec:diff-num-unknown}, discussing the impact of the proportion of pre-knowledge and training samples on recognition performance, as well as the extent to which NDM improves recognition ability in different models.

\subsection{Performance Evaluation of Our M3S-UPD method}
\label{subsec: perfor-eval}
\begin{figure*}
    \centering
    \begin{subfigure}[b]{0.49\textwidth}
        \includegraphics[width=\linewidth]{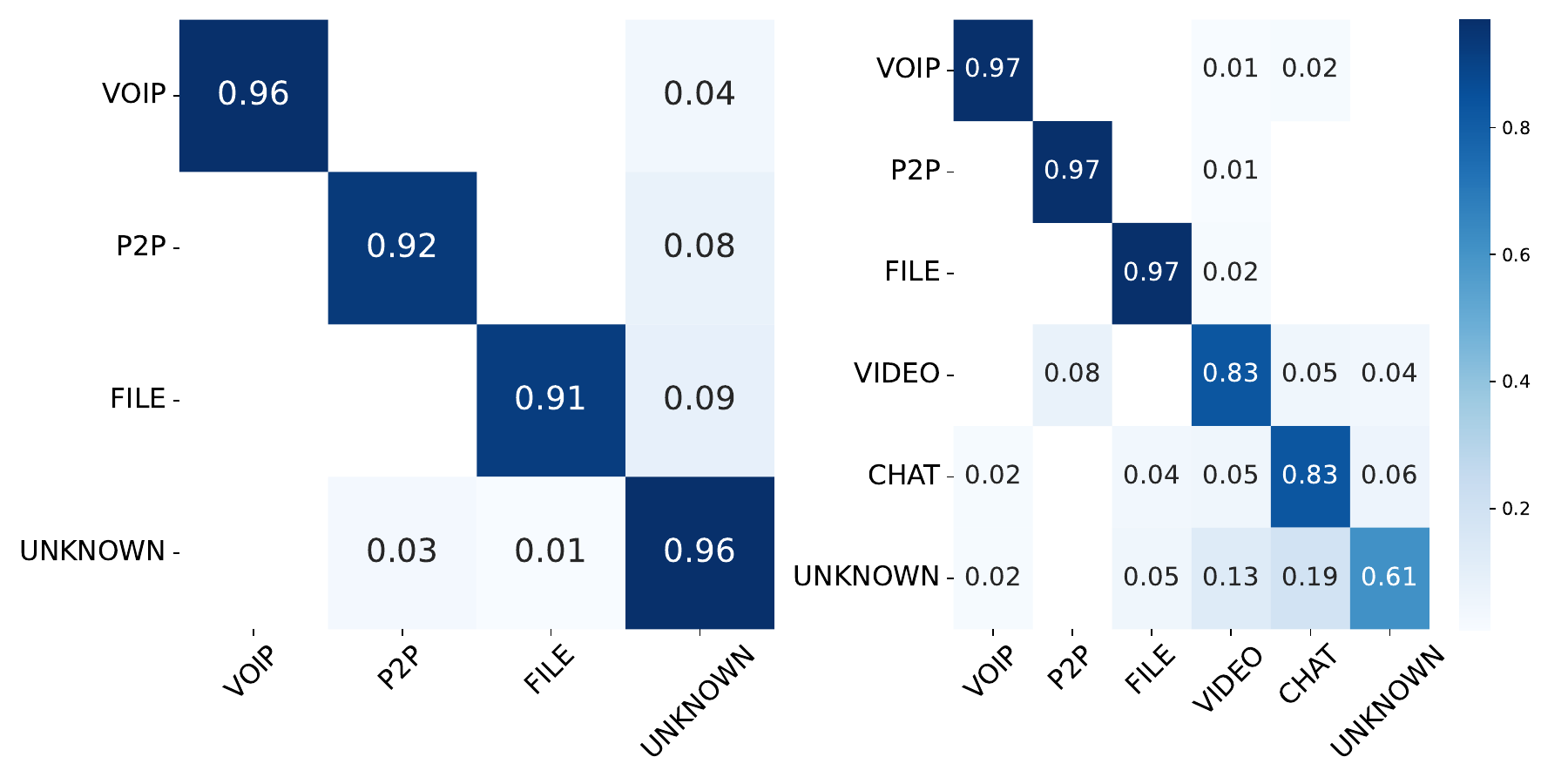}
        \caption{ISCXTor}
        \label{fig: confusion on cictor}
    \end{subfigure}
    \begin{subfigure}[b]{0.49\textwidth}
        \includegraphics[width=\linewidth]{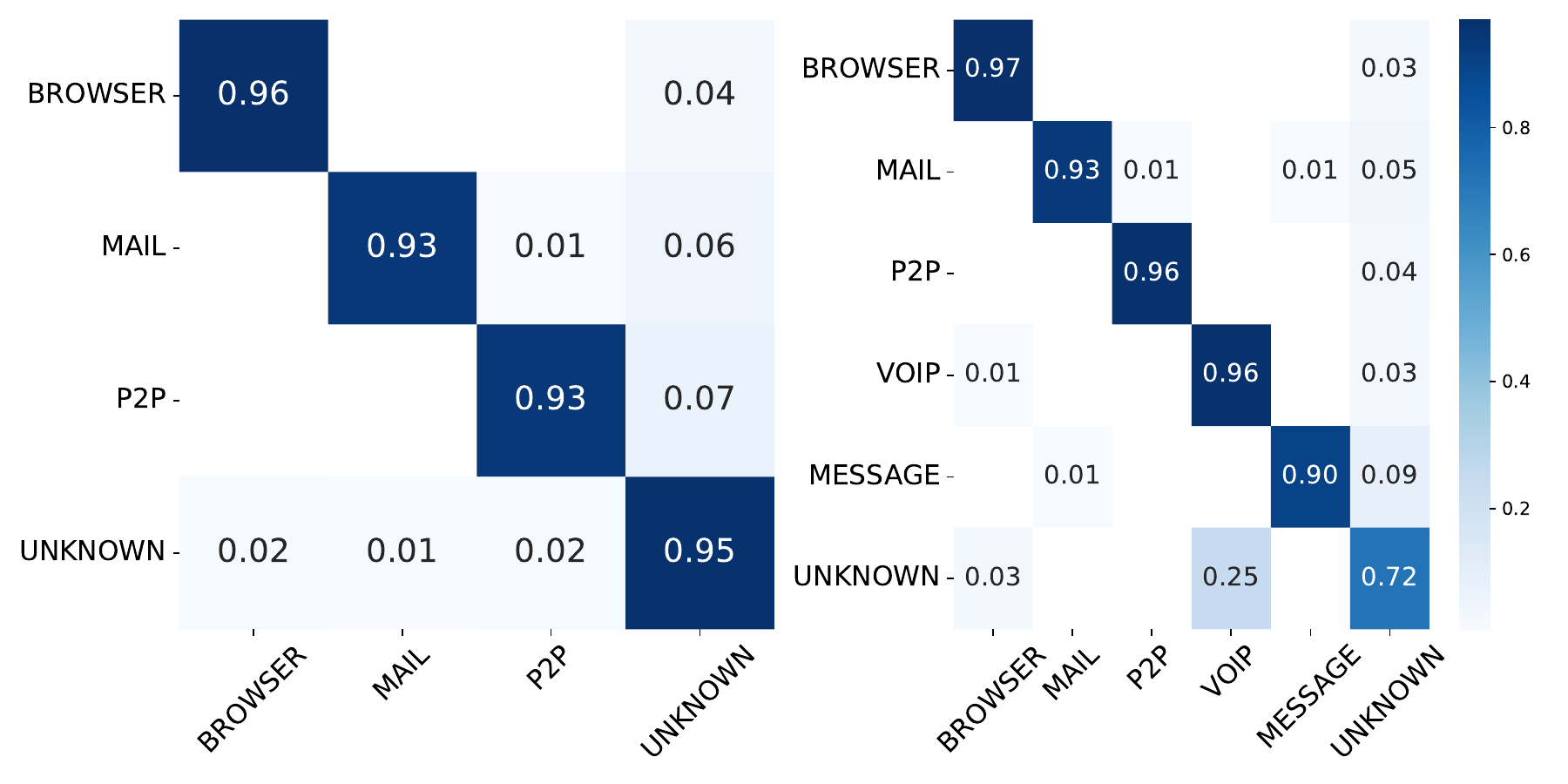}
        \caption{TDTor}
        \label{fig: confusion on tdtor}
    \end{subfigure}
            
    \caption{Confusion matrix in setting1 and setting2 without expert.}
    \label{fig: confusion without expert}
\end{figure*}

We first present the classification results of our proposed M3S-UPD method in four experiment setting, in term of normalized confusion matrix.  Fig.~\ref{fig: confusion without expert} presents the classification results of our method on ISCXTor and TDTor datasets, without the use of expert knowledge. The raws of the confusion matrix indicate the ground truth flow labels, and the columns indicate the predicted labels. The elements on the diagonal of the confusion matrix represent the classification accuracy, while the other elements represent the classification error rate. The darker the color on the diagonal line, the better the classification result. From the diagonal elements in Fig.~\ref{fig: confusion on cictor}, we can see that our method achieves a high classification accuracy of no less than 83\% for all known classes in the absence of expert knowledge on ISCXTor. Also, the classifier recognizes most of the unknown traffic despite the fact that it has never learned it. Similarly, as shown in Fig~\ref{fig: confusion on tdtor}, our method achieving an accuracy of no less than 90\% for all known classes on TDTor. This proves the validity of our spatial distribution alignment process. Although the classifier does not learn any knowledge about the unknown traffic, we greatly ensure the high confidence classification results for the known traffic by evaluating the sample clustering results and the confidence level of model classification results during the consistency check process, thus identifying the unknown traffic. Beyond this, we note that when the number of known classes increases, the model's ability to recognize unknown traffic decreases. This is because the increase in the number of known classes increases the model complexity and therefore the difficulty of consistency checking. Nevertheless, we achieve high classification accuracy for all known classes.

\begin{figure*}
    \centering
    \label{fig: confusion other}
    \begin{subfigure}[b]{0.49\textwidth}
        \includegraphics[width=\linewidth]{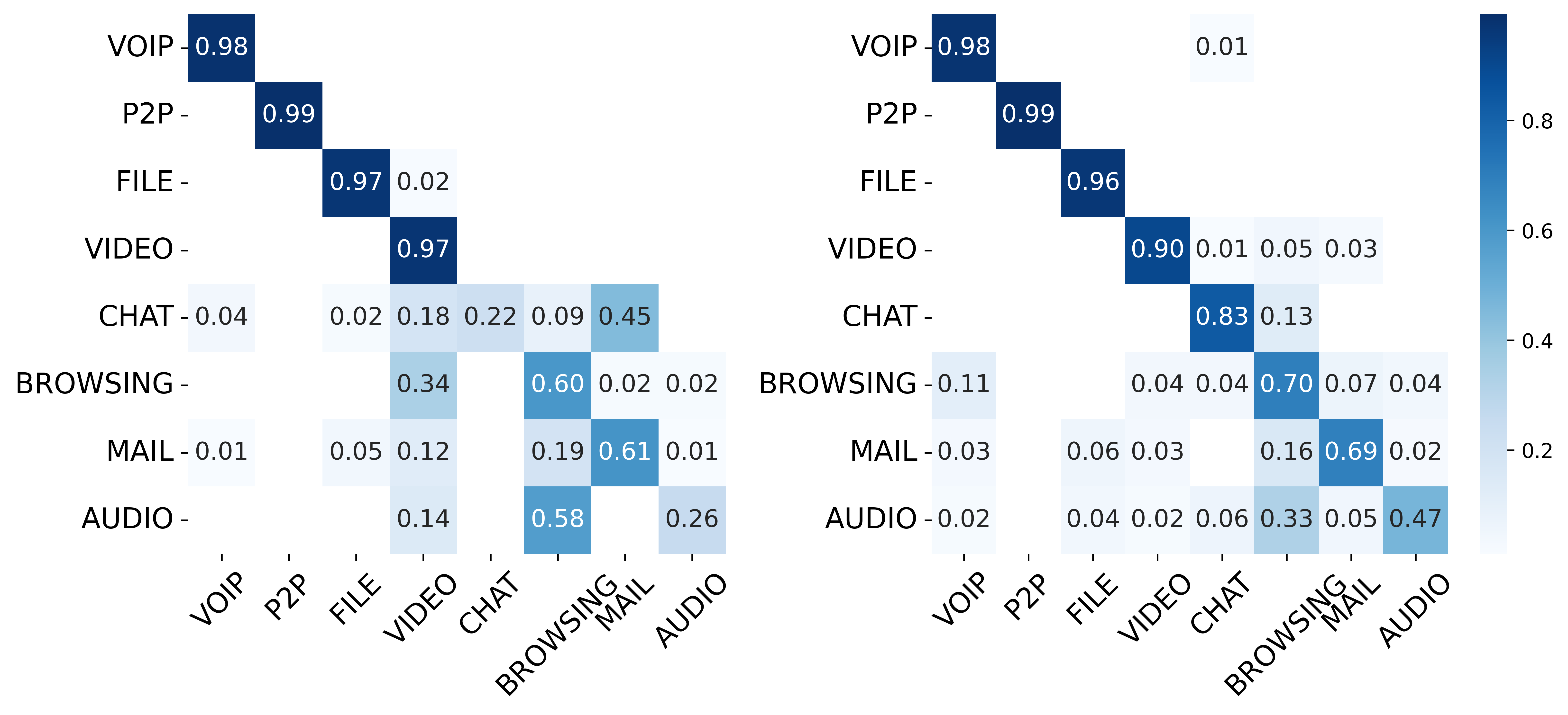}
        \caption{Most classes as known classes}
        \label{fig: Most classes as known classes}
    \end{subfigure}
    \begin{subfigure}[b]{0.49\textwidth}
        \includegraphics[width=\linewidth]{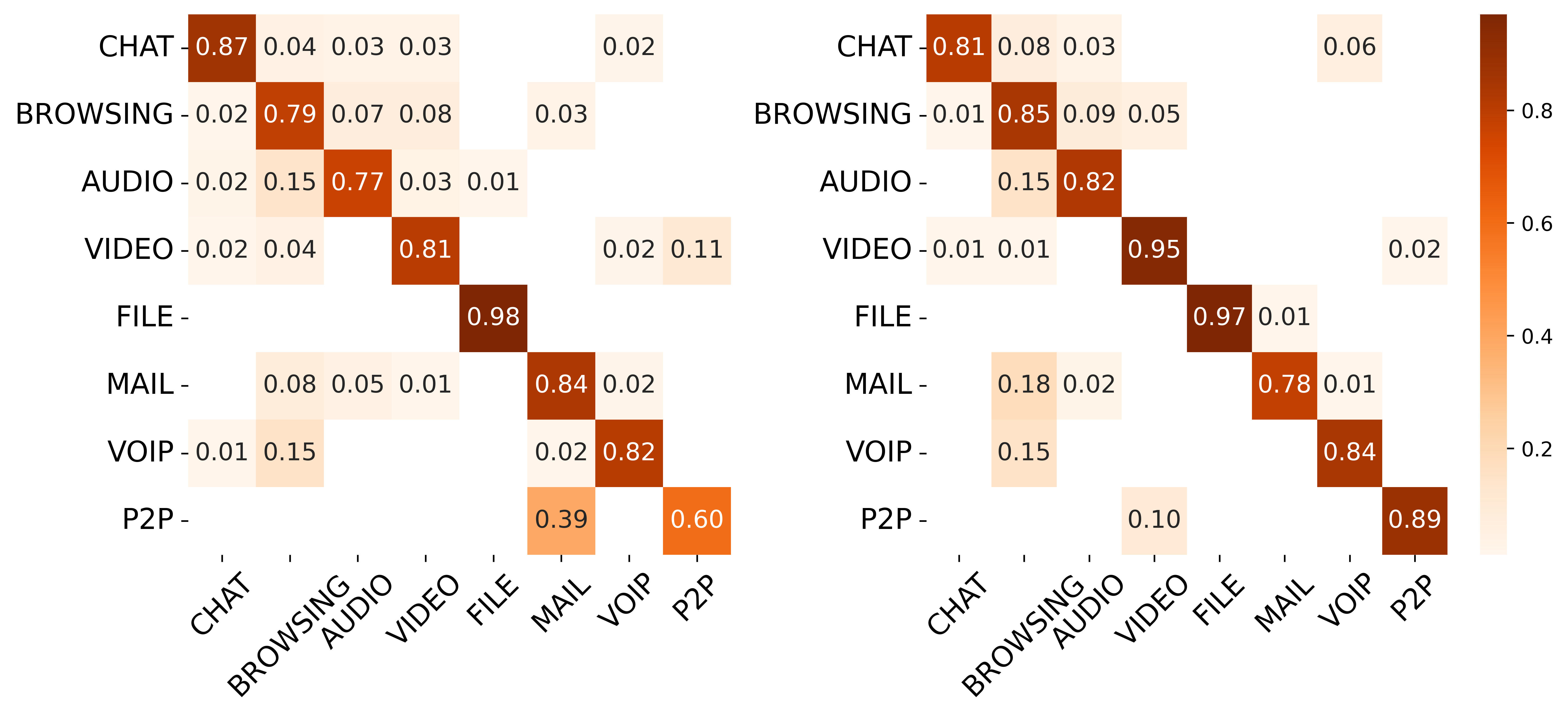}
        \caption{The hardest class to recognize as a known classes}
        \label{fig: The hardest class to recognize as a known classes}
    \end{subfigure}
            
    \caption{Confusion matrix for ICSXTor dataset in setting1 and setting2 with expert for different known classes.}
\end{figure*}

Fig.~\ref{fig: Most classes as known classes} shows the performance of our method for setting1 and setting2 where the number of known classes are 3 and 5, and the samples that fail in consistency check are labeled by introducing the Expert Knowledge on ISCXTor. As can be observed from the left panel (depicted in blue) that all known classes (i.e., those classes with training samples) obtain a high-performance classification of no less than 83\%. In different settings, the model always obtains high accuracy for VoIP, P2P and File-transfer traffic, no less than 96\%. Also, by introducing expert knowledge, the model could recognize all unknown classes. We can see that the accuracy of the add classes is not high although the consistency check process can recognize many unknown classes when the number of known classes is small on ISCXTor. In setting 1, where only VoIP, P2P and File-transfer traffic are involved in initial training, the model successfully learns Video, Browsing and Mail traffic, obtaining accuracy of 97\%, 60\% and 61\%. However, the model seems to easily misclassify Chat as Mail traffic and Audio as Browsing traffic. Things comes different when more classes are known. In setting 2, VoIP, P2P, File Transfer, Video and Chat traffic are involved in the training of the initial model. During the model iteration, M3S-UPD successfully discovers the remaining Browsing, Mail and Audio traffic and by introducing expert knowledge, M3S-UPD correctly labels these traffic and updates the original model with accuracy of 70\%, 69\% and 47\%, which is a substantial improvement compared to setting 1 for all unknown categories. This demonstrates the effectiveness of our proposed M3S-UPD approach for unknown traffic discovery and new class learning. Also, we note that the model in setting 2 hardly misclassifies Chat traffic as Mail traffic due to the knowledge of Chat traffic learned during initial training. At the same time, the results of this experiment to some extent indicate that the initial model has different recognition abilities for different traffic flows, and if it can learn traffic flows of easily confusing classes during the initial training, the M3S-UPD will significantly improve the recognition ability in the subsequent updates.


\begin{figure*}[!htb]
    \centering
    \includegraphics[width=1\linewidth]{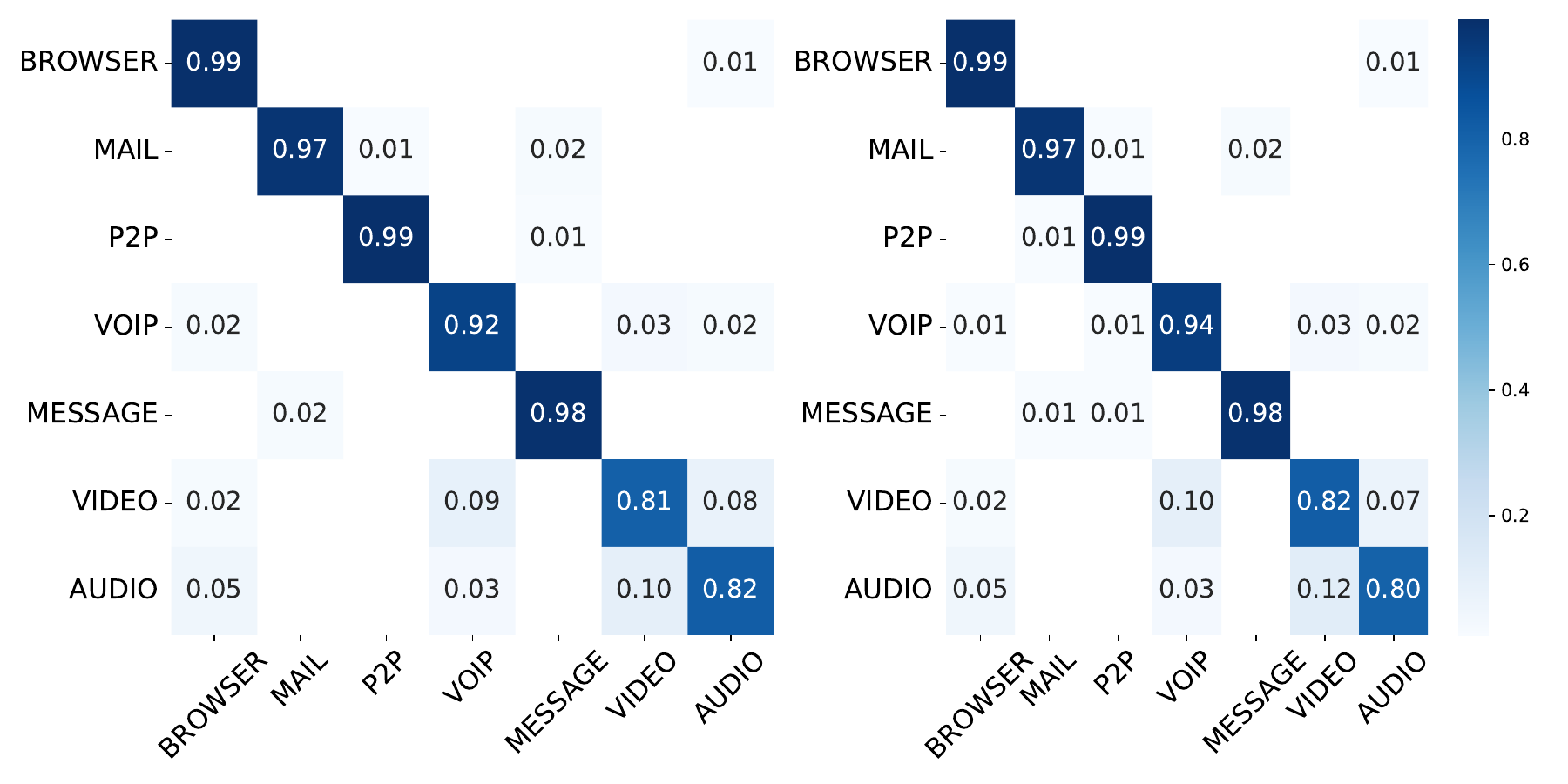}
    \caption{Confusion matrix for TDTor in setting1 and setting2 with expert}
    \label{fig:2.confusion on tdtor with expert}
\end{figure*}

In the experiments detailed within the blue confusion matrices, we consistently selected the class with the largest sample size as the known classes. This approach operates under the assumption that unknown classes are invariably minority classes. Under this known class configuration, our methodology demonstrated high accuracy for known classes but exhibited poor recognition performance for individual unknown classes. It is important to note that the difficulty of recognizing different classes in the original dataset varies, with some classes being inherently more challenging to identify. To mitigate the influence of the known class configuration on recognition outcomes, we redefined the known classes by designating those that are more challenging to recognize as the known classes. The right side (depicted in orange) of Fig~\ref{fig: The hardest class to recognize as a known classes} illustrates the recognition outcomes following the reclassification of known classes. The results indicate a marked improvement in classification accuracy, with the lowest accuracy for P2P traffic not falling below 60\% in setting1, and exceeding 78\% across all categories in setting2. This underscores the significant impact that the order of known classes has on recognition outcomes. When the classifier is pre-trained on classes that are difficult to recognize, the M3S-UPD is sufficiently capable of identifying unknown classes. This highlights the critical role of using known classes to train the classifier. Furthermore, it is noteworthy that P2P traffic, which is the most readily recognized as a known class (with accuracy exceeding 99\%), achieves an accuracy of only 60\% and 89\% when treated as an unknown class in setting1 and setting2, respectively. In contrast, FILE traffic is consistently recognized with very high accuracy (exceeding 97\%) regardless of its status as a known or unknown class. This suggests that FILE traffic is distinctly separable from other traffic types, rendering it easily recognizable as an unknown class and readily learnable.

Fig~\ref{fig:2.confusion on tdtor with expert} illustrates the performance of our method on TDTor under the two settings, with expert knowledge incorporated. It can be observed that our method consistently maintains a high accuracy, with the accuracy for all known classes being no less than 97\% and for all unknown classes no less than 80\%. This contrasts with the situation in ISCXTor, where even when the model has not learned information about easily confusable classes, our method still performs well in distinguishing them when expert knowledge is incorporated. This can be attributed to the more balanced class distribution in TDTor compared to ISCXTor, which suggests that, under a balanced class distribution, our model is capable of accurately identifying each class, even those that are inherently prone to confusion.

\subsection{Performance comparisons with state-of-art benchmarks}
\label{subsec:compare-sota}
To showcase the advanced capabilities of our proposed method, we conducted a comparative analysis against a range of state-of-the-art techniques. Given that these techniques lack consistency checking and the integration of expert knowledge processes, we focused solely on their ability to recognize unknown traffic. Table~\ref{tb: sota-iscxtor} and Table~\ref{tb: sota-tdtor} present the classification results for the ISCXTor and TDTor datasets under Setting 1 and Setting 2, respectively, excluding the introduction of expert knowledge. Our proposed method surpasses the other benchmarks in terms of accuracy, precision, recall, and false positive rate (FPR), achieving an accuracy of 94.69\% in ISCXTor Setting 1, 84.56\% in ISCXTor Setting 2, 94.28\% in TDTor Setting 1, and 91.49\% in TDTor Setting 2, underscoring its efficacy in distinguishing between known and unknown traffic. EVM outperforms the other baselines in Setting 1, achieving a maximum accuracy of 81.87\% on ISCXTor and 72.41\% on TDTor, due to its ability to efficiently learn class boundaries with fewer known categories. Its use of Extreme Value Theory (EVT) allows it to accurately identify known classes and reject unknown ones, enhancing classification performance in a setting with fewer classes.
Notably, our method experiences a pronounced decrease in performance in Setting 2 compared to the CVAE-EVT and Cls-Anomaly methods, while EVM also performs worse overall than both in Setting 2. This discrepancy may be attributed to the CVAE-EVT and Cls-Anomaly utilize a two-stage hierarchical detection framework designed to enhance overall recognition performance by minimizing the false alarm rate of benign traffic, which is more effective in Setting 2 where there are more known categories and a larger sample size, thereby enhancing overall accuracy. Nonetheless, it is noteworthy that our method demonstrates superior recognition of unknown traffic compared to all other methods.
\begin{table*}[!htb]
\caption{Comparison of different methods for different setting on ISCXTor.}
\label{tb: sota-iscxtor}
\centering
\begin{tabular}{|c|cccc|cccc|}
\hline
                    & \multicolumn{4}{c|}{setting1   no expert}                                                                                            & \multicolumn{4}{c|}{setting2   no expert}                                                                                            \\ \hline
                    & \multicolumn{1}{c|}{accuracy}        & \multicolumn{1}{c|}{precision}       & \multicolumn{1}{c|}{recall}          & FPR             & \multicolumn{1}{c|}{accuracy}        & \multicolumn{1}{c|}{precision}       & \multicolumn{1}{c|}{recall}          & FPR             \\ \hline
CVAE-EVT            & \multicolumn{1}{c|}{0.7381}          & \multicolumn{1}{c|}{0.7030}          & \multicolumn{1}{c|}{0.5665}          & 0.1009          & \multicolumn{1}{c|}{0.7650}          & \multicolumn{1}{c|}{0.6365}          & \multicolumn{1}{c|}{0.6079}          & 0.0512          \\ \hline
Cls-Anomaly         & \multicolumn{1}{c|}{0.7991}          & \multicolumn{1}{c|}{0.7790}          & \multicolumn{1}{c|}{0.8256}          & 0.0645          & \multicolumn{1}{c|}{0.7791}          & \multicolumn{1}{c|}{0.6878}          & \multicolumn{1}{c|}{0.6443}          & 0.0476          \\ \hline
EVM                 & \multicolumn{1}{c|}{0.8187}          & \multicolumn{1}{c|}{0.8145}          & \multicolumn{1}{c|}{0.8046}          & 0.0673          & \multicolumn{1}{c|}{0.7733}          & \multicolumn{1}{c|}{0.7064}          & \multicolumn{1}{c|}{0.7710}          & 0.0432          \\ \hline
\textbf{Our Method} & \multicolumn{1}{c|}{\textbf{0.9469}} & \multicolumn{1}{c|}{\textbf{0.9480}} & \multicolumn{1}{c|}{\textbf{0.9365}} & \textbf{0.0204} & \multicolumn{1}{c|}{\textbf{0.8456}} & \multicolumn{1}{c|}{\textbf{0.7812}} & \multicolumn{1}{c|}{\textbf{0.8619}} & \textbf{0.0289} \\ \hline
\end{tabular}
\end{table*}

\begin{table*}[!htb]
\caption{Comparison of different methods for different setting on TDTor.}
\label{tb: sota-tdtor}
\centering
\begin{tabular}{|c|cccc|cccc|}
\hline
                    & \multicolumn{4}{c|}{setting1   no expert}                                                                                            & \multicolumn{4}{c|}{setting2   no expert}                                                                                            \\ \hline
                    & \multicolumn{1}{c|}{accuracy}        & \multicolumn{1}{c|}{precision}       & \multicolumn{1}{c|}{recall}          & FPR             & \multicolumn{1}{c|}{accuracy}        & \multicolumn{1}{c|}{precision}       & \multicolumn{1}{c|}{recall}          & FPR             \\ \hline
CVAE-EVT            & \multicolumn{1}{c|}{0.6903}          & \multicolumn{1}{c|}{0.7152}          & \multicolumn{1}{c|}{0.7416}          & 0.1100          & \multicolumn{1}{c|}{0.8187}          & \multicolumn{1}{c|}{0.7009}          & \multicolumn{1}{c|}{0.7901}          & 0.0367          \\ \hline
Cls-Anomaly         & \multicolumn{1}{c|}{0.7175}          & \multicolumn{1}{c|}{0.7475}          & \multicolumn{1}{c|}{0.7912}          & 0.0961          & \multicolumn{1}{c|}{0.8008}          & \multicolumn{1}{c|}{0.8119}          & \multicolumn{1}{c|}{0.7842}          & 0.0395          \\ \hline
EVM                 & \multicolumn{1}{c|}{0.7241}          & \multicolumn{1}{c|}{0.7794}          & \multicolumn{1}{c|}{0.6697}          & 0.1158          & \multicolumn{1}{c|}{0.7442}          & \multicolumn{1}{c|}{0.6951}          & \multicolumn{1}{c|}{0.7211}          & 0.0517          \\ \hline
\textbf{Our Method} & \multicolumn{1}{c|}{\textbf{0.9428}} & \multicolumn{1}{c|}{\textbf{0.9471}} & \multicolumn{1}{c|}{\textbf{0.9409}} & \textbf{0.0222} & \multicolumn{1}{c|}{\textbf{0.9149}} & \multicolumn{1}{c|}{\textbf{0.9146}} & \multicolumn{1}{c|}{\textbf{0.9067}} & \textbf{0.0169} \\ \hline
\end{tabular}
\end{table*}

\begin{figure*}
    \centering
    \caption{Classification Results of different methods when varying the proportion of known classes in different settings.}
    \label{fig: Classification Results of different methods when varying the proportion of known classes in different settings.}
    \begin{subfigure}[b]{1\textwidth}
        \includegraphics[width=\linewidth]{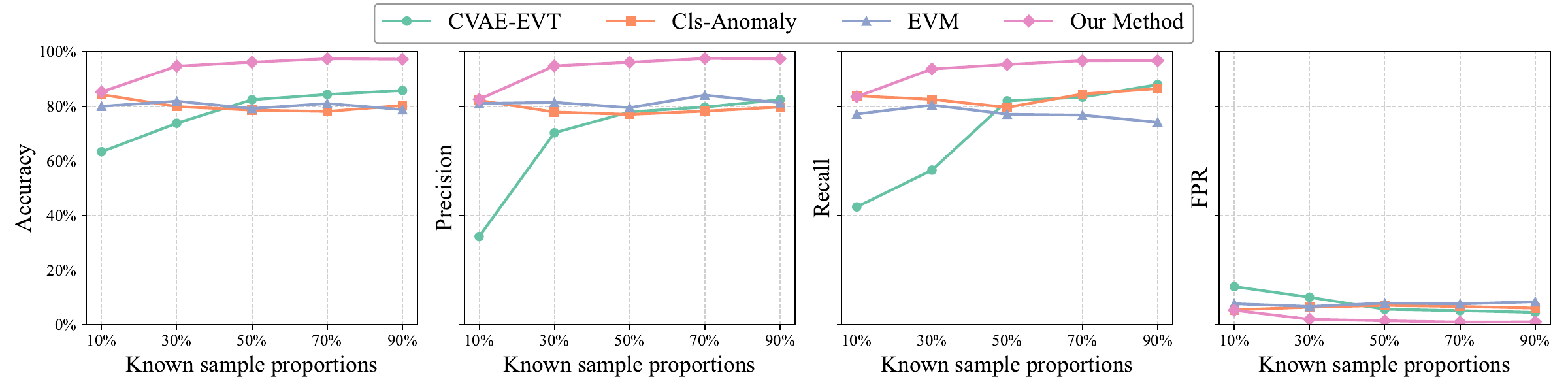}
        \caption{ISCXTor setting1}
        \label{fig: CICTor setting1}
    \end{subfigure}
    \begin{subfigure}[b]{1\textwidth}
        \includegraphics[width=\linewidth]{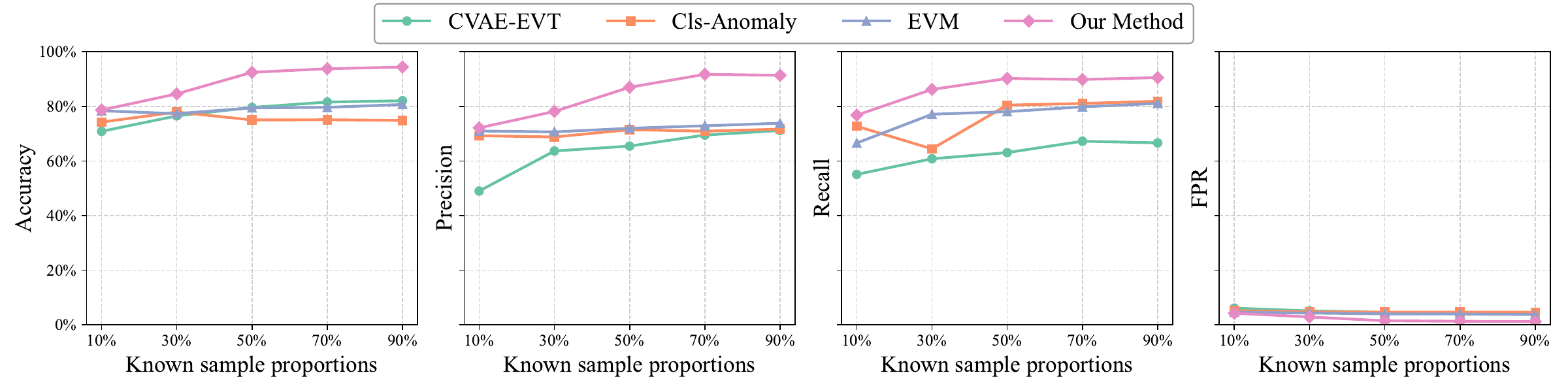}
        \caption{ISCXTor setting2}
        \label{fig: CICTor setting2}
    \end{subfigure}
    \begin{subfigure}[b]{1\textwidth}
        \includegraphics[width=\linewidth]{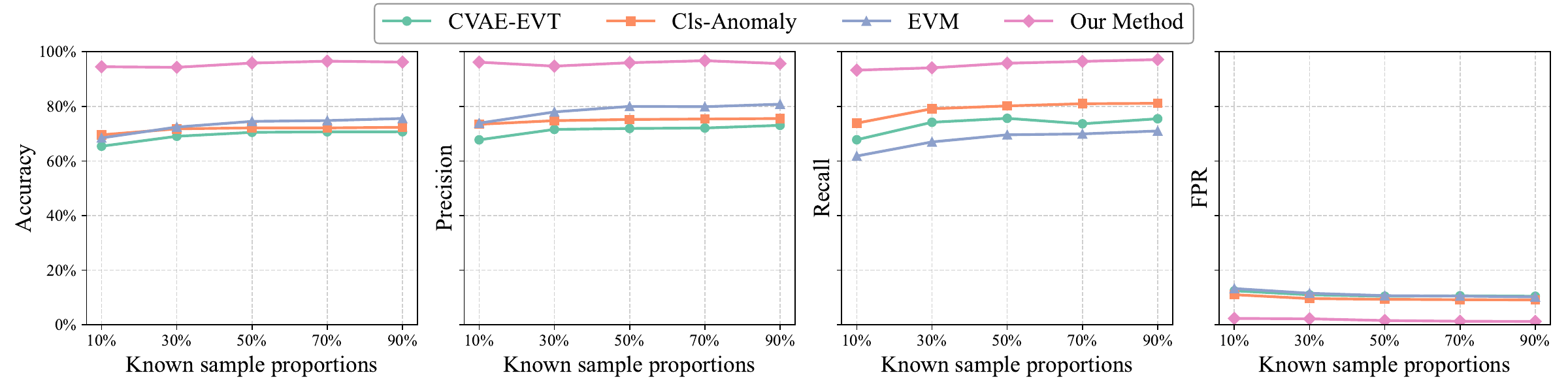}
        \caption{TDTor setting1}
        \label{fig: TDTor setting1}
    \end{subfigure}
    \begin{subfigure}[b]{1\textwidth}
        \includegraphics[width=\linewidth]{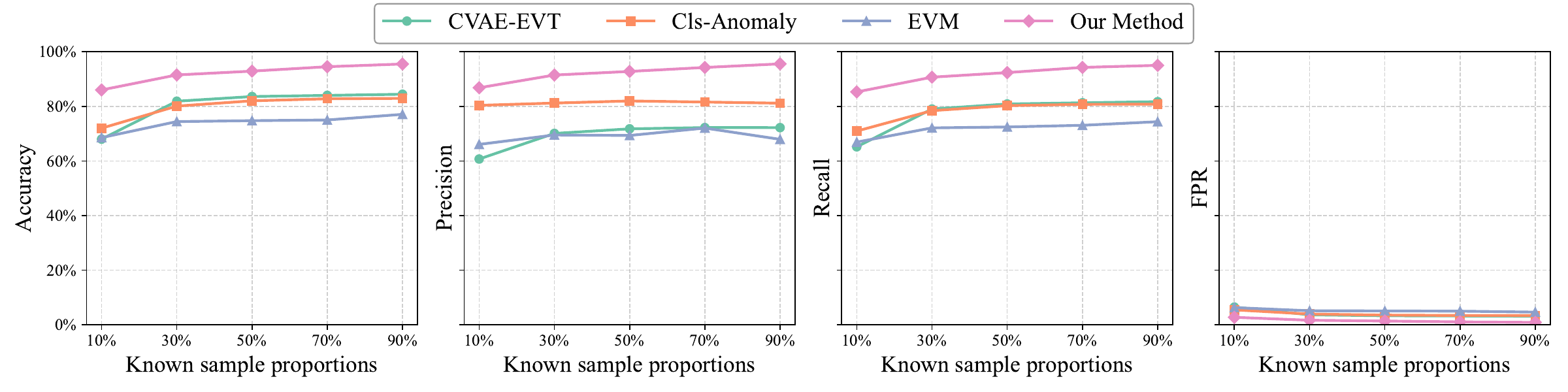}
        \caption{TDTor setting2}
        \label{fig: TDTor setting2}
    \end{subfigure}
        
\end{figure*}

To further investigate the traffic classification capabilities of our method compared to these state-of-the-art techniques across various scenarios, we present in Fig~\ref{fig: Classification Results of different methods when varying the proportion of known classes in different settings.} the classification performance of these methods with varying proportions of training samples from known classes in different settings across two datasets. Our findings reveal that the proposed M3S-UPD method consistently outperforms the other methods in both accuracy and precision of unknown traffic as the proportion of known samples increases. In all four figures, it is evident that our method consistently achieves superior classification performance compared to CVAE-EVT, Cls-Anomaly, and EVM when utilizing more than 10\% of the known sample proportion. This outcome substantiates the effectiveness of our method in identifying unknown traffic.

Moreover, classifier performance exhibits a gradual improvement as the proportion of known samples increases. This trend can be attributed to the fact that a higher proportion of known samples enables the initial classifier to acquire more comprehensive knowledge of the samples. Additionally, we observe that in Figures~\ref{fig: CICTor setting1} and~\ref{fig: TDTor setting1}, 30\% of the training samples suffice for the classifier to attain stable classification performance. Conversely, in Figures~\ref{fig: CICTor setting2} and~\ref{fig: TDTor setting2}, 50\% of the training samples are necessary for the classifier's performance to approach its maximum potential. This observation suggests that when there are more original training categories, a greater number of initial samples is required for the original classifier to acquire sufficient classification knowledge.

\subsection{Classification Performance with Different Number and proportion of Known Classes}
\label{sub:diff-num-known}

The experiment conducted has highlighted the efficacy of our methodology in discerning unidentified network traffic even with a limited knowledge base. To evaluate how the volume of pre-existing knowledge influences the performance of our method, we examined the effects of varying the number of known and unknown classes used for training on the identification results.

Fig~\ref{fig:Accuracy, FPR and proportion of expert knowledge of varying number of known classes on ISCXTor.} illustrates the identification outcomes, including metrics such as Accuracy, False Positive Rate (FPR), and the proportion of expert knowledge introduced, across different quantities of known class types, while maintaining a constant training set size of 30\% within our comprehensive framework. Similarly, Fig~\ref{fig:Accuracy, FPR and proportion of expert knowledge of varying number of known classes on TDTor.} shows the corresponding identification outcomes on TDTor under the same conditions. A notable increase in the accuracy of the Network Discovery Method (M3S-UPD) is observed as the number of known classes increases. In the "Without Expert Knowledge" scenario depicted in Fig~\ref{fig:Accuracy, FPR and proportion of expert knowledge of varying number of known classes on ISCXTor.}, when utilizing a single known class, M3S-UPD achieves an accuracy of 95.99\%. In contrast, when employing seven known classes, thereby reducing the scenario to a single unknown class, M3S-UPD attains an accuracy of approximately 83.29\%. Similarly, in Fig~\ref{fig:Accuracy, FPR and proportion of expert knowledge of varying number of known classes on TDTor.}, when utilizing a single known class, M3S-UPD achieves an accuracy of 98.13\%, when employing six known classes, the accuracy is 90.54\%. As the number of known classes increases, the accuracy decreases because more known classes result in more classification tasks and decision boundaries. This increased complexity makes it more challenging for the model to distinguish between categories. However, as the number of known classes increases, the accuracy of our method does not significantly decrease, this demonstrates M3S-UPD's capability to accurately recognize known classes.

\begin{figure}
    \centering
    \includegraphics[width=\linewidth]{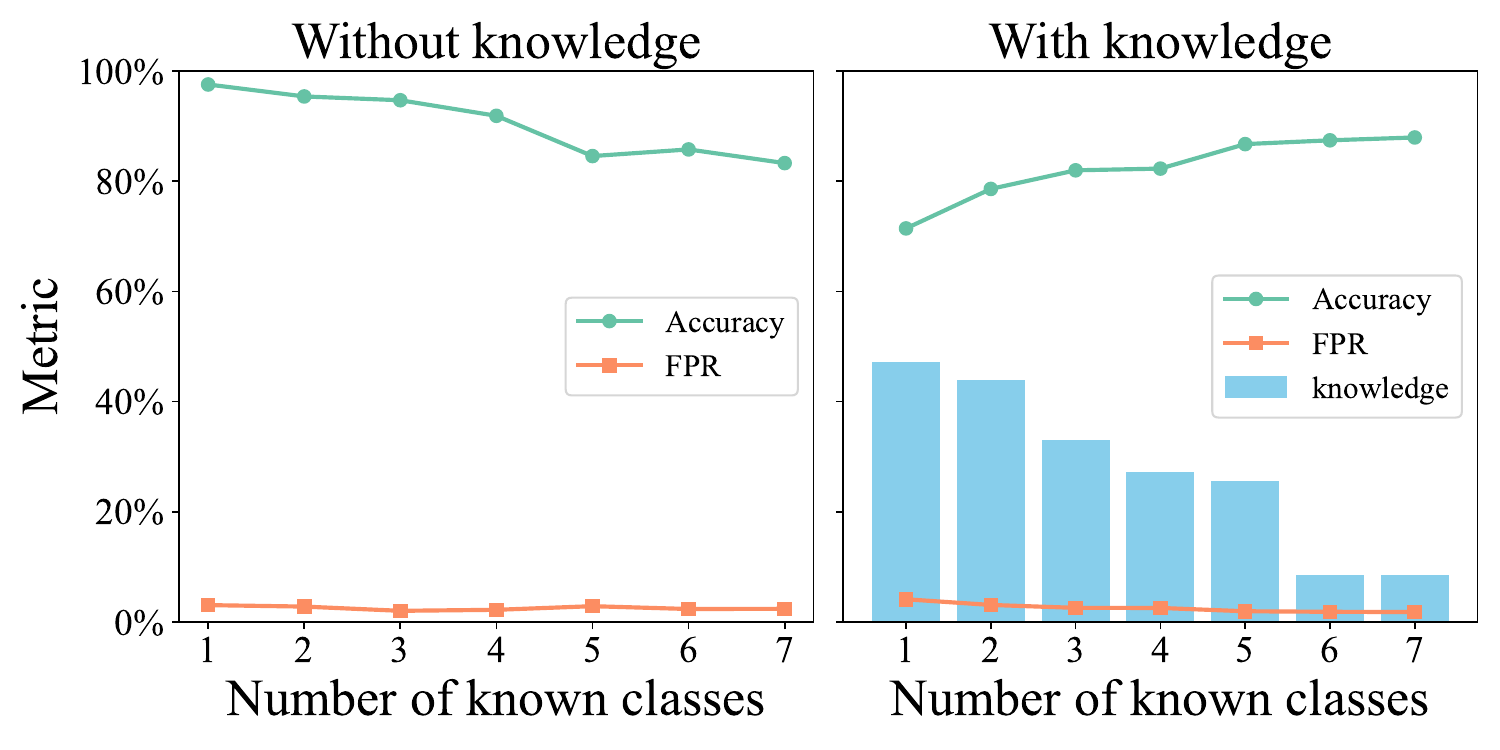}
    \caption{Accuracy, FPR and proportion of expert knowledge of varying number of known classes on ISCXTor.}
    \label{fig:Accuracy, FPR and proportion of expert knowledge of varying number of known classes on ISCXTor.}
\end{figure}

\begin{figure}
    \centering
    \includegraphics[width=\linewidth]{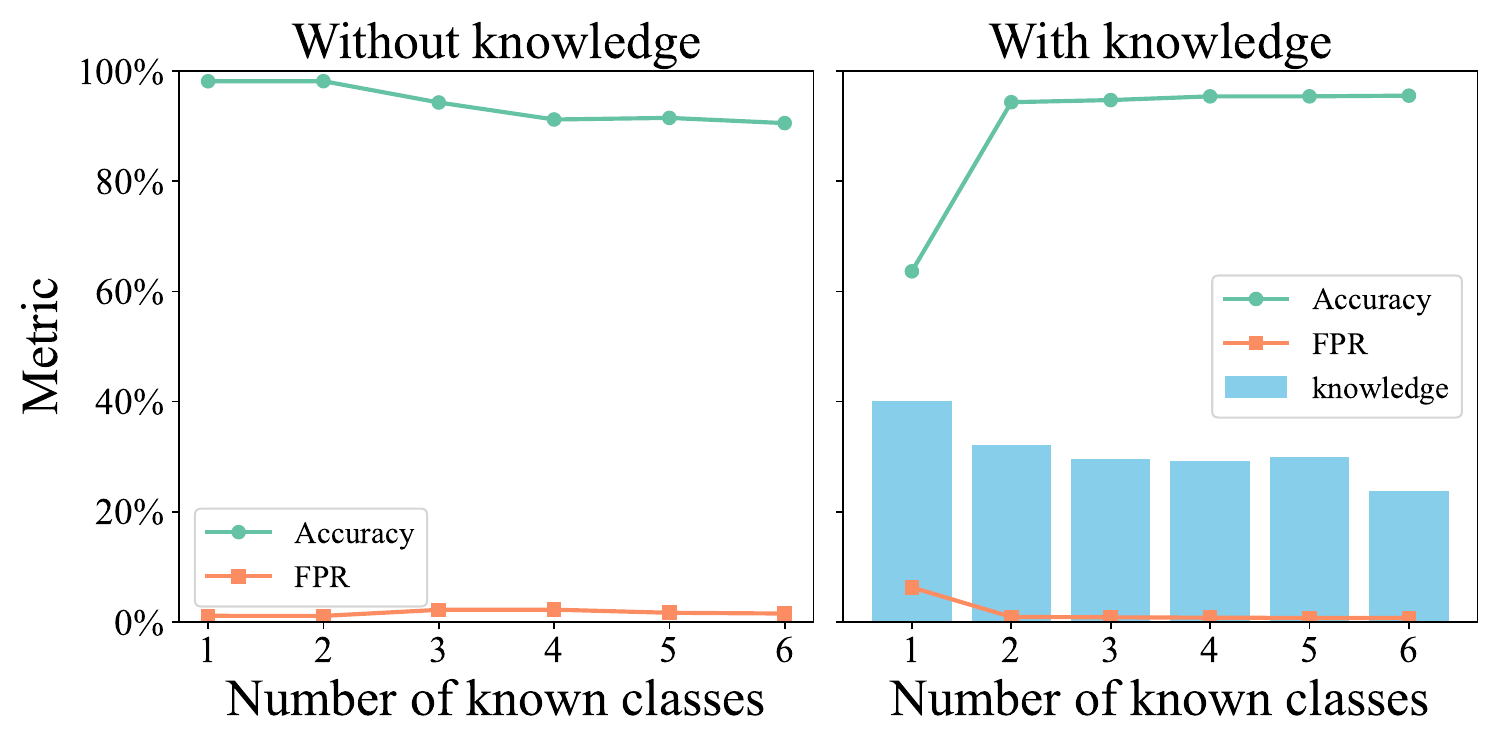}
    \caption{Accuracy, FPR and proportion of expert knowledge of varying number of known classes on TDTor.}
    \label{fig:Accuracy, FPR and proportion of expert knowledge of varying number of known classes on TDTor.}
\end{figure}

In the "With Expert Knowledge" scenario depicted in Fig~\ref{fig:Accuracy, FPR and proportion of expert knowledge of varying number of known classes on ISCXTor.}, M3S-UPD achieves an accuracy of 71.43\% on the ICSXTor, necessitating 47.24\% of expert knowledge when the number of known classes is one. When employing seven known classes, M3S-UPD registers an accuracy of 87.93\%, with only 8.58\% of expert knowledge required for the unclassified categories. As shown in Fig~\ref{fig:Accuracy, FPR and proportion of expert knowledge of varying number of known classes on TDTor.}, M3S-UPD demonstrates similar performance on the TDTor dataset, achieving an accuracy of 63.65\% when the number of known classes is one, requiring 40.14\% of expert knowledge. When the number of known classes is six, the accuracy increases to 95.52\%, with only 23.82\% of expert knowledge needed. This demonstrates M3S-UPD's proficiency in seamlessly integrating expert knowledge while simultaneously acquiring profound insights across various categories, ultimately leading to the accurate classification of samples. An upward trajectory in accuracy and precision is observed alongside a reduction in the need for expert knowledge as the number of known categories increases. This phenomenon can be attributed to the model's adeptness in assimilating traffic knowledge from the identified categories. Notably, when three known categories are used for initial model training, M3S-UPD achieves an accuracy of 81.97\%  on ICSXTor, with the proportion of expert knowledge required decreasing from over 47\% to 33\%. Similarly, on the TDTor, M3S-UPD achieves an accuracy of 94.72\%, with the proportion of expert knowledge decreasing from 40.14\% to 29.62\%. Subsequently, a marginal decline in the necessity for expert knowledge is observed as the number of known categories increases. This suggests that a minimum of three known categories is required for M3S-UPD to effectively harvest traffic knowledge from the identified categories.

\begin{figure*}
    \centering
    \caption{Accuracy and FPR of varying proportion of known classes in different settings.}
    \label{fig: Accuracy and FPR of varying proportion of known classes in different settings.}
    \begin{subfigure}[b]{1\textwidth}
        \includegraphics[width=\linewidth]{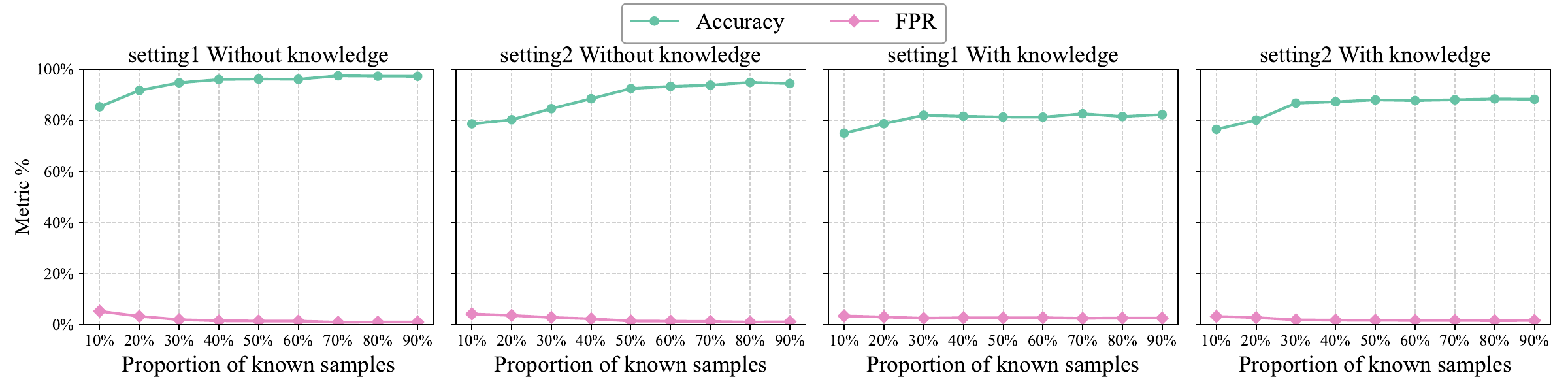}
        \caption{ISCXTor}
        \label{fig: CICTor knwown samole proportions}
    \end{subfigure}
    \begin{subfigure}[b]{1\textwidth}
        \includegraphics[width=\linewidth]{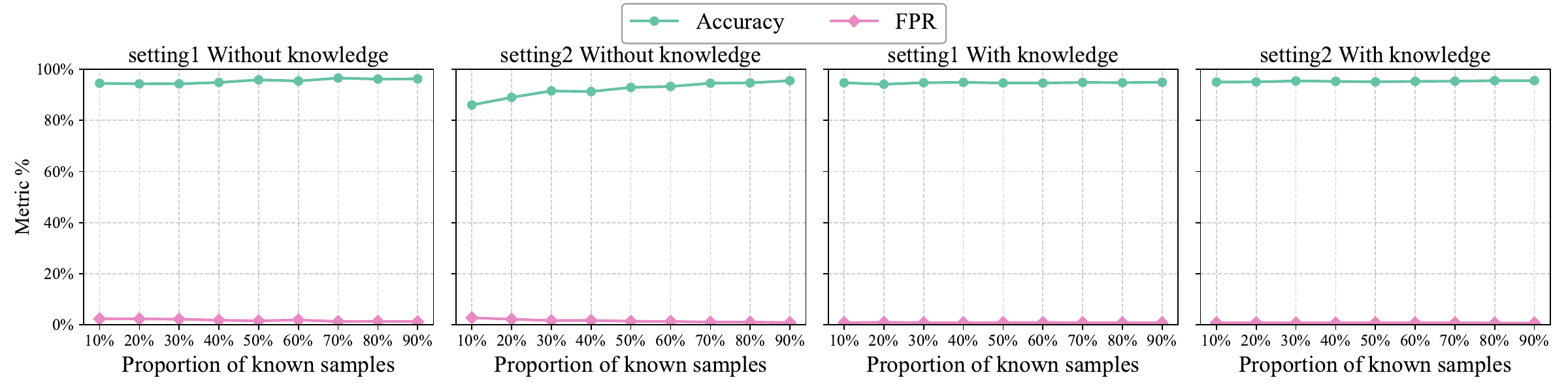}
        \caption{TDTor}
        \label{fig: TDTor knwown samole proportions}
    \end{subfigure}
        
\end{figure*}

We aimed to evaluate the impact of the number of samples used in training the initial model on the final classification performance of the classifier. Fig~\ref{fig: CICTor knwown samole proportions} and Fig~\ref{fig: TDTor knwown samole proportions} illustrates the accuracy and False Positive Rate (FPR) as the proportion of samples from the initial known classes is increased across different settings for ISCXTor and TDTor, respectively. It is evident that accuracy gradually improves with an increase in training samples across various settings. Without the introduction of expert knowledge, the accuracy of Setting 1 and Setting 2 rises from 85.28\% to 97.24\% and from 78.63\% to 94.38\%, respectively, as the percentage of known samples increases from 10\% to 90\%. Similarly, on the TDTor dataset, the accuracy of Setting 1 increases from 94.49\% to 96.20\%, and the accuracy of Setting 2 rises from 85.9\% to 95.49\% as the percentage of known samples increases from 10\% to 90\%. With the inclusion of expert knowledge, the accuracies for Setting 1 and Setting 2 are 81.97\% and 86.73\%, respectively, when using a 30\% proportion of known samples. On the TDTor dataset, the accuracies for Setting 1 and Setting 2 are 94.69\% and 94.98\%, respectively, when using a 10\% proportion of known samples. Beyond this point, increasing the proportion of training samples does not lead to a significant improvement in accuracy. This indicates that our proposed M3S-UPD method can effectively extract knowledge from existing samples and accurately identify unknown classes. Additionally, by incorporating expert knowledge, our method can label uncertain samples, thereby achieving stable traffic classification.

\begin{figure*}
    \centering
    \caption{Accuracy and FPR of varying the number of unknown classes for different settings.}
    \label{fig: Accuracy and FPR of varying the number of unknown classes for different settings.}
    \begin{subfigure}[b]{1\textwidth}
        \includegraphics[width=\linewidth]{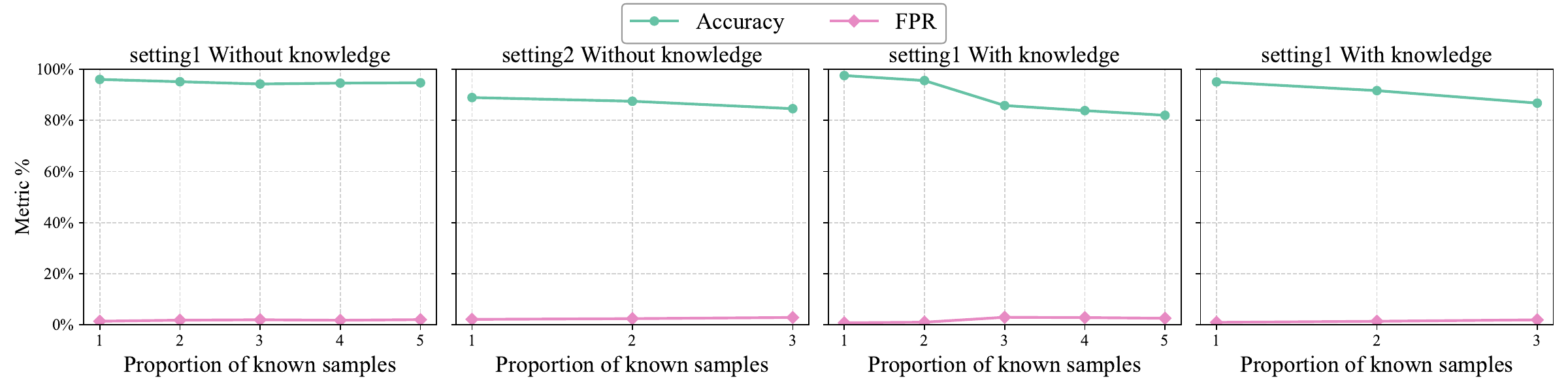}
        \caption{ISCXTor}
        \label{fig: CICTor unknwown samole proportions}
    \end{subfigure}
    \begin{subfigure}[b]{1\textwidth}
        \includegraphics[width=\linewidth]{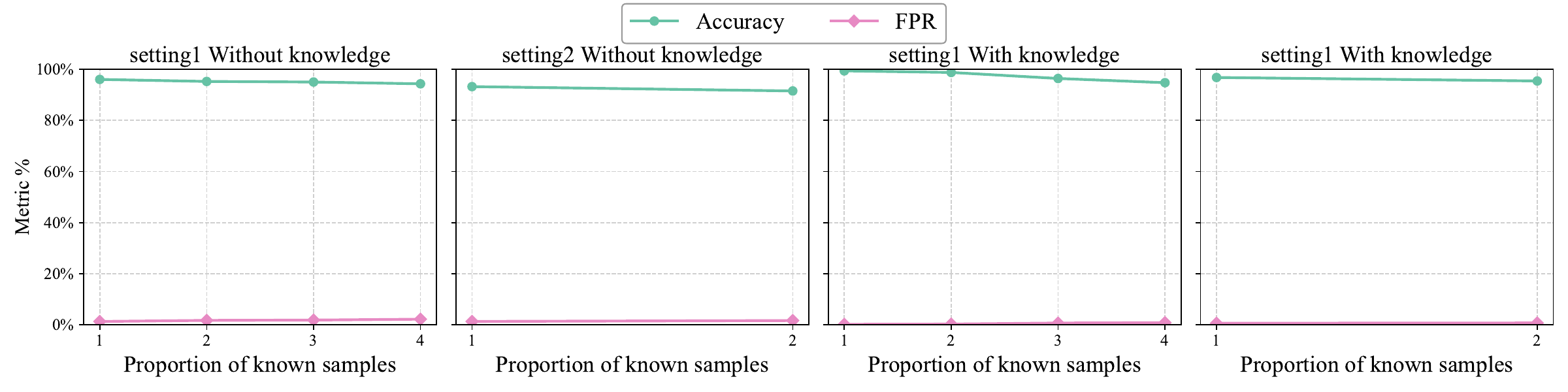}
        \caption{TDTor}
        \label{fig: TDTor unknwown samole proportions}
    \end{subfigure}
        
\end{figure*}

\subsection{Classification Performance with Different Number of Unknown Classes}
\label{subsec:diff-num-unknown}
We sought to assess the impact of varying numbers of unknown classes on categorization. Fig~\ref{fig: Accuracy and FPR of varying the number of unknown classes for different settings.} displays the accuracy and False Positive Rate (FPR) as the number of unknown classes changes across different settings for two datasets. To ensure the effectiveness of the initial training, we use 3 known classes for both datasets and 30\% known samples for each known class to train the initial classifiers. It is evident that with a fixed number of known classes, increasing the number of unknown classes results in a gradual decline in overall accuracy. This phenomenon occurs in both datasets. For ISCXTor dataset, when the number of known classes is three, and no expert knowledge is introduced, increasing the number of unknown classes from 1 to 5 leads to a slight decrease in accuracy from 95.99\% to 94.69\%. In contrast, when expert knowledge is incorporated, the accuracy decreases from 97.51\% to 81.97\%, indicating a noticeable drop. For TDTor dataset, the accuracy decrease from 96\% to 94.28\% and 99.35\% to 94.72\%, separately. 
We observe that even with an increase in the number of unknown classes, there is only a slight decrease in overall recognition accuracy. This phenomenon likely results from the enriched knowledge base to which the initial model is exposed, allowing for more precise judgments on unknown categories. When expert knowledge is introduced to achieve fine-grained classification, the increase in the number of unknown classes results in a slightly more pronounced decrease in accuracy due to the heightened recognition granularity. This underscores M3S-UPD's ability to effectively integrate expert knowledge in identifying unknown network traffic, with the precision of the initial training data being a crucial determinant of the quality of the final identification outcomes.

\section{Conclusion}
\label{Conclusion}

This paper presents M3S-UPD, a novel self-supervised training framework designed to address the challenges of encrypted traffic classification and unknown traffic detection under limited labeled data conditions. By leveraging unlabeled traffic data through iterative model refinement, our framework incrementally improves classification performance while simultaneously identifying unknown traffic without relying on prior knowledge or synthetic data augmentation. The key innovation lies in the integration of embedding clustering, spatial distribution alignment, and consistency-based pseudo-labeling, which enables the model to distinguish between known and unknown traffic categories effectively. Experimental results on public datasets demonstrate that M3S-UPD achieves competitive performance compared to state-of-the-art methods in both closed-world and open-world scenarios. The framework’s ability to adapt to concept drifting and continuously update its knowledge base makes it particularly suitable for real-world network environments where traffic patterns evolve dynamically.
The success of M3S-UPD highlights several important directions for future research in encrypted traffic analysis. Firstly, the potential integration of more advanced clustering techniques could enhance the framework’s ability to discern subtle differences between traffic categories. Besides, the development of more efficient model update mechanisms could further reduce the computational overhead associated with continuous learning in real-time network environments.


%



\section*{Acknowledgment}
This work was supported in part by the National Key Research and Development Program of China (Grant No. 2023YFB3106700) under the Young Scientists Program,  in part by Natural Science Foundation of Jiangsu Province  (Grant No. SBK2023041256), and in part by the National Natural Science Foundation of China (Grant No. 62302097).

\ifCLASSOPTIONcaptionsoff
  \newpage
\fi



%



\bibliographystyle{IEEEtran}
\bibliography{reference}

\begin{thebibliography}{10}
\providecommand{\url}[1]{#1}
\csname url@samestyle\endcsname
\providecommand{\newblock}{\relax}
\providecommand{\bibinfo}[2]{#2}
\providecommand{\BIBentrySTDinterwordspacing}{\spaceskip=0pt\relax}
\providecommand{\BIBentryALTinterwordstretchfactor}{4}
\providecommand{\BIBentryALTinterwordspacing}{\spaceskip=\fontdimen2\font plus
\BIBentryALTinterwordstretchfactor\fontdimen3\font minus \fontdimen4\font\relax}
\providecommand{\BIBforeignlanguage}[2]{{%
\expandafter\ifx\csname l@#1\endcsname\relax
\typeout{** WARNING: IEEEtran.bst: No hyphenation pattern has been}%
\typeout{** loaded for the language `#1'. Using the pattern for}%
\typeout{** the default language instead.}%
\else
\language=\csname l@#1\endcsname
\fi
#2}}
\providecommand{\BIBdecl}{\relax}
\BIBdecl

\bibitem{Barradas2021FlowLensEE}
\BIBentryALTinterwordspacing
D.~Barradas, N.~Santos, L.~Rodrigues, S.~Signorello, F.~M.~V. Ramos, and A.~Madeira, ``Flowlens: Enabling efficient flow classification for ml-based network security applications,'' \emph{Proceedings 2021 Network and Distributed System Security Symposium}, 2021. [Online]. Available: \url{https://api.semanticscholar.org/CorpusID:231590874}
\BIBentrySTDinterwordspacing

\bibitem{Yun2023EncryptedTT}
\BIBentryALTinterwordspacing
X.~chun Yun, Y.~Wang, Y.~Zhang, C.~Zhao, and Z.~Zhao, ``Encrypted tls traffic classification on cloud platforms,'' \emph{IEEE/ACM Transactions on Networking}, vol.~31, pp. 164--177, 2023. [Online]. Available: \url{https://api.semanticscholar.org/CorpusID:250705780}
\BIBentrySTDinterwordspacing

\bibitem{Wang2023ATA}
Y.~Wang, H.~S. He, Y.~K. Lai, and A.~X. Liu, ``A two-phase approach to fast and accurate classification of encrypted traffic,'' \emph{IEEE/ACM Transactions on Networking}, vol.~31, pp. 1071--1086, 2023.

\bibitem{Li2022PacketLevelOA}
\BIBentryALTinterwordspacing
J.~Li, S.~Wu, H.~Zhou, X.~Luo, T.~Wang, Y.~Liu, and X.~Ma, ``Packet-level open-world app fingerprinting on wireless traffic,'' \emph{Proceedings 2022 Network and Distributed System Security Symposium}, 2022. [Online]. Available: \url{https://api.semanticscholar.org/CorpusID:248224842}
\BIBentrySTDinterwordspacing

\bibitem{Fu2023FrequencyDF}
\BIBentryALTinterwordspacing
C.~Fu, Q.~Li, M.~Shen, and K.~Xu, ``Frequency domain feature based robust malicious traffic detection,'' \emph{IEEE/ACM Transactions on Networking}, vol.~31, pp. 452--467, 2023. [Online]. Available: \url{https://api.semanticscholar.org/CorpusID:251459863}
\BIBentrySTDinterwordspacing

\bibitem{Bai2008TrafficIO}
\BIBentryALTinterwordspacing
X.~Bai, Y.~Zhang, and X.~Niu, ``Traffic identification of tor and web-mix,'' \emph{2008 Eighth International Conference on Intelligent Systems Design and Applications}, vol.~1, pp. 548--551, 2008. [Online]. Available: \url{https://api.semanticscholar.org/CorpusID:11519694}
\BIBentrySTDinterwordspacing

\bibitem{Sicker2007LegalIS}
\BIBentryALTinterwordspacing
D.~C. Sicker, P.~Ohm, and D.~Grunwald, ``Legal issues surrounding monitoring during network research,'' in \emph{ACM/SIGCOMM Internet Measurement Conference}, 2007. [Online]. Available: \url{https://api.semanticscholar.org/CorpusID:17998633}
\BIBentrySTDinterwordspacing

\bibitem{Finsterbusch2014ASO}
\BIBentryALTinterwordspacing
M.~Finsterbusch, C.~Richter, E.~Rocha, J.-A. M{\"u}ller, and K.~Hanssgen, ``A survey of payload-based traffic classification approaches,'' \emph{IEEE Communications Surveys \& Tutorials}, vol.~16, pp. 1135--1156, 2014. [Online]. Available: \url{https://api.semanticscholar.org/CorpusID:20792087}
\BIBentrySTDinterwordspacing

\bibitem{AlSabahBG12}
M.~AlSabah, K.~S. Bauer, and I.~Goldberg, ``Enhancing tor's performance using real-time traffic classification,'' in \emph{the {ACM} Conference on Computer and Communications Security, CCS'12, Raleigh, NC, USA, October 16-18, 2012}, T.~Yu, G.~Danezis, and V.~D. Gligor, Eds.\hskip 1em plus 0.5em minus 0.4em\relax {ACM}, 2012, pp. 73--84.

\bibitem{Cuzzocrea2017TorTA}
A.~Cuzzocrea, F.~Martinelli, F.~Mercaldo, and G.~V. Vercelli, ``Tor traffic analysis and detection via machine learning techniques,'' \emph{2017 IEEE International Conference on Big Data (Big Data)}, pp. 4474--4480, 2017.

\bibitem{Montieri_Anonymity}
A.~Montieri, D.~Ciuonzo, G.~Aceto, and A.~Pescapé, ``Anonymity services tor, i2p, jondonym: Classifying in the dark (web),'' \emph{IEEE Transactions on Dependable and Secure Computing}, vol.~17, no.~3, pp. 662--675, 2020.

\bibitem{PathSignature22}
S.~Xu, G.~Geng, X.~Jin, D.~Liu, and J.~Weng, ``Seeing traffic paths: Encrypted traffic classification with path signature features,'' \emph{{IEEE} Trans. Inf. Forensics Secur.}, vol.~17, pp. 2166--2181, 2022.

\bibitem{WFatIS16}
A.~Panchenko, F.~Lanze, J.~Pennekamp, T.~Engel, A.~Zinnen, M.~Henze, and K.~Wehrle, ``Website fingerprinting at internet scale,'' in \emph{23rd Annual Network and Distributed System Security Symposium, {NDSS} 2016, San Diego, California, USA, February 21-24, 2016}.\hskip 1em plus 0.5em minus 0.4em\relax The Internet Society, 2016.

\bibitem{Cherubin2022OnlineWF}
G.~Cherubin, R.~Jansen, and C.~Troncoso, ``Online website fingerprinting: Evaluating website fingerprinting attacks on tor in the real world,'' in \emph{USENIX Security Symposium}, 2022.

\bibitem{Liu_FSNet}
C.~Liu, L.~He, G.~Xiong, Z.~Cao, and Z.~Li, ``Fs-net: {A} flow sequence network for encrypted traffic classification,'' in \emph{2019 {IEEE} Conference on Computer Communications, {INFOCOM} 2019, Paris, France, April 29 - May 2, 2019}.\hskip 1em plus 0.5em minus 0.4em\relax {IEEE}, 2019, pp. 1171--1179.

\bibitem{Zhao2022FlowSA}
\BIBentryALTinterwordspacing
R.~Zhao, X.~Deng, Y.~Wang, L.~Chen, M.~Liu, Z.~Xue, and Y.~Wang, ``Flow sequence-based anonymity network traffic identification with residual graph convolutional networks,'' \emph{2022 IEEE/ACM 30th International Symposium on Quality of Service (IWQoS)}, pp. 1--10, 2022. [Online]. Available: \url{https://api.semanticscholar.org/CorpusID:250317945}
\BIBentrySTDinterwordspacing

\bibitem{Zhao2023YetAT}
\BIBentryALTinterwordspacing
R.~Zhao, M.~Zhan, X.~Deng, Y.~Wang, Y.~Wang, G.~Gui, and Z.~Xue, ``Yet another traffic classifier: A masked autoencoder based traffic transformer with multi-level flow representation,'' in \emph{AAAI Conference on Artificial Intelligence}, 2023. [Online]. Available: \url{https://api.semanticscholar.org/CorpusID:259716837}
\BIBentrySTDinterwordspacing

\bibitem{Deng2023RobustMW}
\BIBentryALTinterwordspacing
X.~Deng, Q.~Yin, Z.~Liu, X.~Zhao, Q.~Li, M.~Xu, K.~Xu, and J.~Wu, ``Robust multi-tab website fingerprinting attacks in the wild,'' \emph{2023 IEEE Symposium on Security and Privacy (SP)}, pp. 1005--1022, 2023. [Online]. Available: \url{https://api.semanticscholar.org/CorpusID:260002990}
\BIBentrySTDinterwordspacing

\bibitem{snWF22}
Y.~Wang, H.~Xu, Z.~Guo, Z.~Qin, and K.~Ren, ``snwf: Website fingerprinting attack by ensembling the snapshot of deep learning,'' \emph{{IEEE} Trans. Inf. Forensics Secur.}, vol.~17, pp. 1214--1226, 2022.

\bibitem{Ede2020FlowPrintSM}
\BIBentryALTinterwordspacing
T.~van Ede, R.~Bortolameotti, A.~Continella, J.~Ren, D.~J. Dubois, M.~Lindorfer, D.~R. Choffnes, M.~van Steen, and A.~Peter, ``Flowprint: Semi-supervised mobile-app fingerprinting on encrypted network traffic,'' \emph{Proceedings 2020 Network and Distributed System Security Symposium}, 2020. [Online]. Available: \url{https://api.semanticscholar.org/CorpusID:211265114}
\BIBentrySTDinterwordspacing

\bibitem{Fu2023DetectingUE}
\BIBentryALTinterwordspacing
C.~Fu, Q.~Li, and K.~Xu, ``Detecting unknown encrypted malicious traffic in real time via flow interaction graph analysis,'' \emph{ArXiv}, vol. abs/2301.13686, 2023. [Online]. Available: \url{https://api.semanticscholar.org/CorpusID:256415981}
\BIBentrySTDinterwordspacing

\bibitem{Attarian2019AdaWFPAAO}
\BIBentryALTinterwordspacing
R.~Attarian, L.~Abdi, and S.~Hashemi, ``Adawfpa: Adaptive online website fingerprinting attack for tor anonymous network: A stream-wise paradigm,'' \emph{Comput. Commun.}, vol. 148, pp. 74--85, 2019. [Online]. Available: \url{https://api.semanticscholar.org/CorpusID:203704356}
\BIBentrySTDinterwordspacing

\bibitem{Zhang2020AutonomousUF}
\BIBentryALTinterwordspacing
J.~Zhang, F.~Li, F.~Ye, and H.~Wu, ``Autonomous unknown-application filtering and labeling for dl-based traffic classifier update,'' \emph{IEEE INFOCOM 2020 - IEEE Conference on Computer Communications}, pp. 397--405, 2020. [Online]. Available: \url{https://api.semanticscholar.org/CorpusID:211132760}
\BIBentrySTDinterwordspacing

\bibitem{Wang2016OnRA}
\BIBentryALTinterwordspacing
T.~Wang and I.~Goldberg, ``On realistically attacking tor with website fingerprinting,'' \emph{Proceedings on Privacy Enhancing Technologies}, vol. 2016, pp. 21 -- 36, 2016. [Online]. Available: \url{https://api.semanticscholar.org/CorpusID:26413501}
\BIBentrySTDinterwordspacing

\bibitem{Payap2019TFMP}
\BIBentryALTinterwordspacing
P.~Sirinam, N.~Mathews, M.~S. Rahman, and M.~Wright, ``Triplet fingerprinting: More practical and portable website fingerprinting with n-shot learning,'' in \emph{Proceedings of the 2019 ACM SIGSAC Conference on Computer and Communications Security}, ser. CCS '19.\hskip 1em plus 0.5em minus 0.4em\relax New York, NY, USA: Association for Computing Machinery, 2019, p. 1131–1148. [Online]. Available: \url{https://doi.org/10.1145/3319535.3354217}
\BIBentrySTDinterwordspacing

\bibitem{Oh2021GANDaLFGF}
\BIBentryALTinterwordspacing
S.~E. Oh, N.~Mathews, M.~S. Rahman, M.~K. Wright, and N.~Hopper, ``Gandalf: Gan for data-limited fingerprinting,'' \emph{Proceedings on Privacy Enhancing Technologies}, vol. 2021, pp. 305 -- 322, 2021. [Online]. Available: \url{https://api.semanticscholar.org/CorpusID:231779670}
\BIBentrySTDinterwordspacing

\bibitem{Zhou2023FewshotWF}
\BIBentryALTinterwordspacing
Q.~Zhou, L.~Wang, H.~Zhu, and T.~Lu, ``Few-shot website fingerprinting attack with cluster adaptation,'' \emph{Comput. Networks}, vol. 229, p. 109780, 2023. [Online]. Available: \url{https://api.semanticscholar.org/CorpusID:258297158}
\BIBentrySTDinterwordspacing

\bibitem{Hu2022AttributeBasedZL}
\BIBentryALTinterwordspacing
Y.~Hu, G.~Cheng, W.~Chen, and B.~Jiang, ``Attribute-based zero-shot learning for encrypted traffic classification,'' \emph{IEEE Transactions on Network and Service Management}, vol.~19, pp. 4583--4599, 2022. [Online]. Available: \url{https://api.semanticscholar.org/CorpusID:249717931}
\BIBentrySTDinterwordspacing

\bibitem{lashkari2017characterization}
A.~H. Lashkari, G.~D. Gil, M.~S.~I. Mamun, and A.~A. Ghorbani, ``Characterization of tor traffic using time based features,'' in \emph{International Conference on Information Systems Security and Privacy}, vol.~2.\hskip 1em plus 0.5em minus 0.4em\relax SciTePress, 2017, pp. 253--262.

\bibitem{chaabane2010digging}
A.~Chaabane, P.~Manils, and M.~A. Kaafar, ``Digging into anonymous traffic: A deep analysis of the tor anonymizing network,'' in \emph{2010 fourth international conference on network and system security}.\hskip 1em plus 0.5em minus 0.4em\relax IEEE, 2010, pp. 167--174.

\bibitem{Yang2021ConditionalVA}
J.~Yang, X.~Chen, S.~Chen, X.~Jiang, and X.~Tan, ``Conditional variational auto-encoder and extreme value theory aided two-stage learning approach for intelligent fine-grained known/unknown intrusion detection,'' \emph{IEEE Transactions on Information Forensics and Security}, vol.~16, pp. 3538--3553, 2021.

\bibitem{baseline2henrydoss2017incremental}
J.~Henrydoss, S.~Cruz, E.~M. Rudd, M.~Gunther, and T.~E. Boult, ``Incremental open set intrusion recognition using extreme value machine,'' in \emph{2017 16th IEEE International Conference on Machine Learning and Applications (ICMLA)}.\hskip 1em plus 0.5em minus 0.4em\relax IEEE, 2017, pp. 1089--1093.

\end{thebibliography}

%

\begin{IEEEbiography}[{\includegraphics[width=1in,height=1.25in,clip,keepaspectratio]{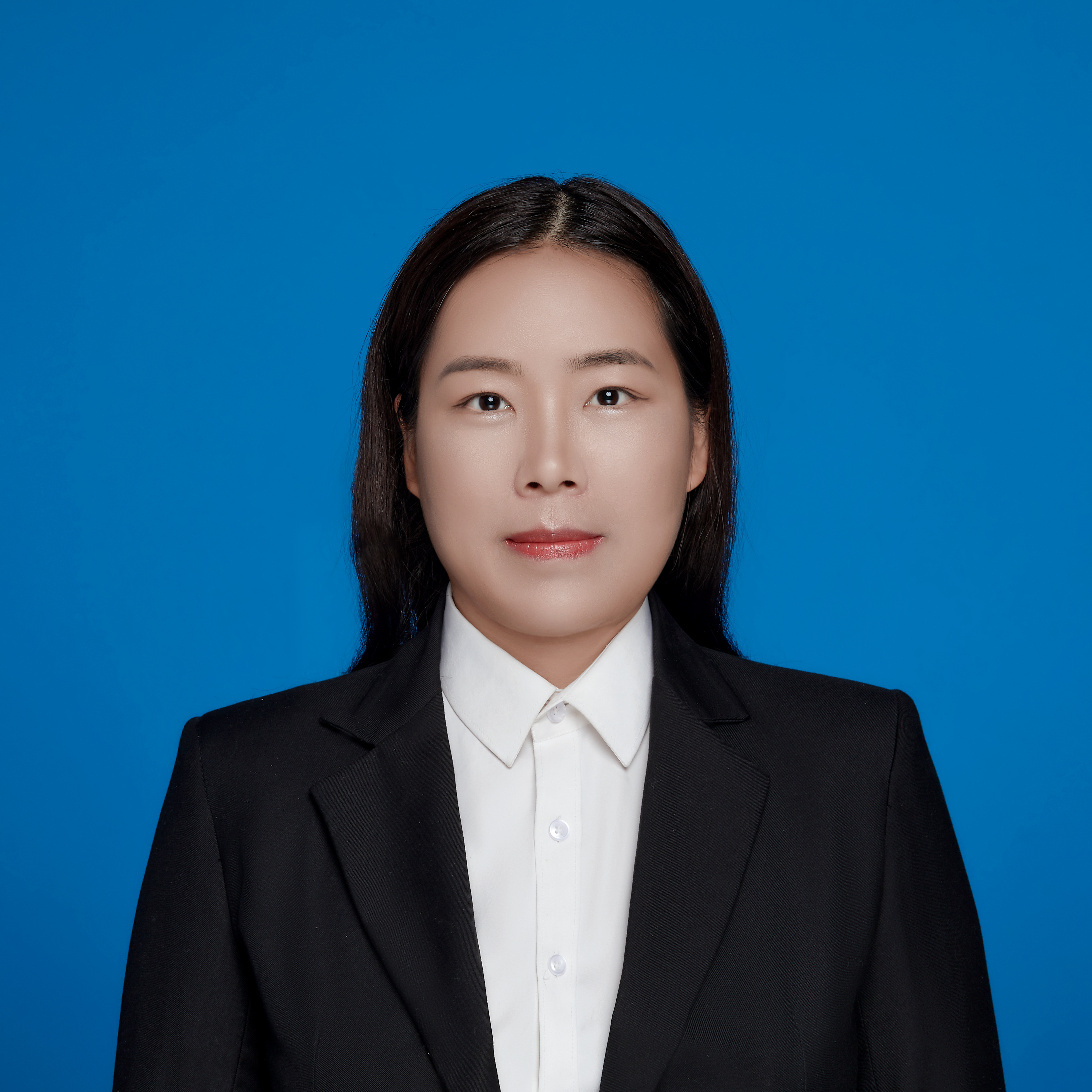}}]{Yali Yuan}
received her Ph.D. degree from Göttingen University (Göttingen, Germany) in 2018. Dr. Yuan joined the School of Cyber Science and Engineering, Southeast University (Nanjing, China), as an associate professor in 2021. Her research interests include network traffic security analysis, network attack and defense, as well as privacy protection.
\end{IEEEbiography}

\vspace{4pt}

\begin{IEEEbiography}[{\includegraphics[width=1in,height=1.25in,clip,keepaspectratio]{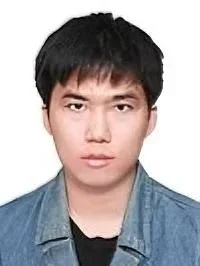}}]{Yu Huang}
 is currently an intern at the School of Cyber Science and Engineering, Southeast University, Nanjing, China. His current research interests include fuzz testing, traffic analysis, and static code analysis.
\end{IEEEbiography}

\vspace{4pt}

\begin{IEEEbiography}[{\includegraphics[width=1in,height=1.25in,clip,keepaspectratio]{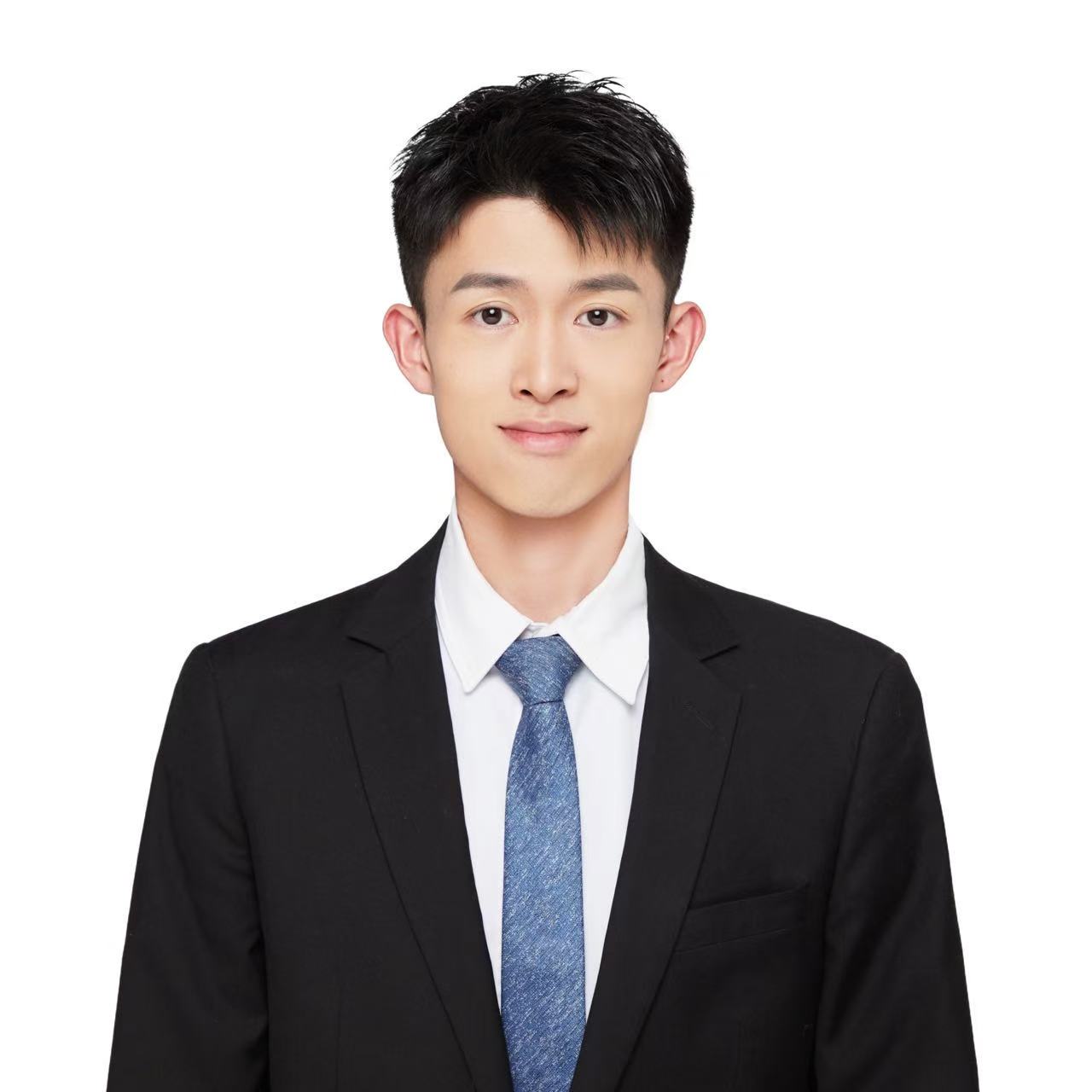}}]{Xingjian Zeng}
 (Student Member, IEEE) is currently pursuing the B.S. degree with the School of Cyber Science and Engineering, Southeast University, Nanjing, China. His current research interests include information security, traffic analysis, and data mining.
\end{IEEEbiography}

\vspace{4pt}

\begin{IEEEbiography}[{\includegraphics[width=1in,height=1.25in,clip,keepaspectratio]{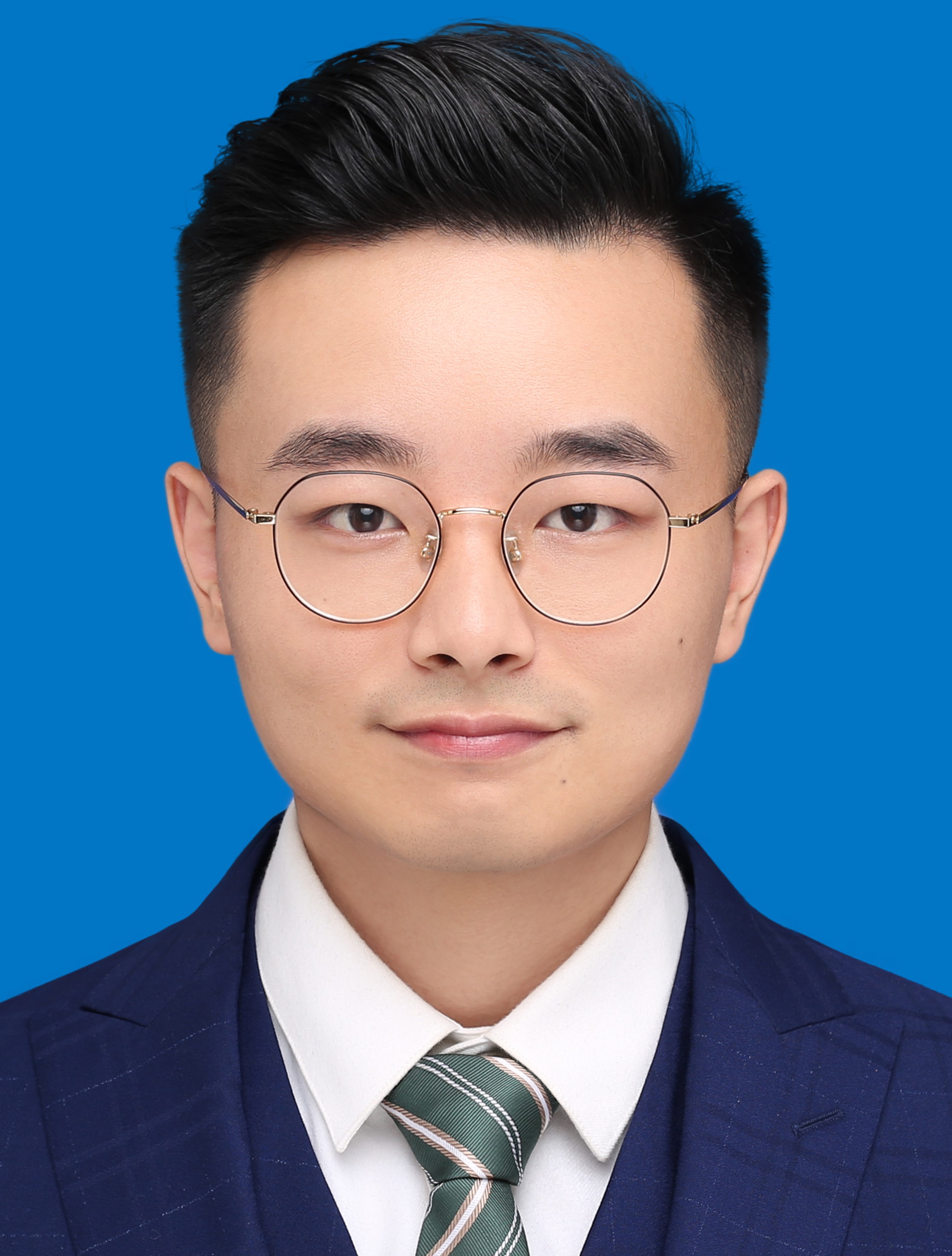}}]{Hantao Mei}
is currently pursuing his Ph.D. degree with the School of Cyber Science and Engineering, Southeast University, Nanjing, China. His current research interests include anonymous communication, traffic analysis, and network measurement.
\end{IEEEbiography}

\vspace{4pt}

\begin{IEEEbiography}[{\includegraphics[width=1in,height=1.25in,clip,keepaspectratio]{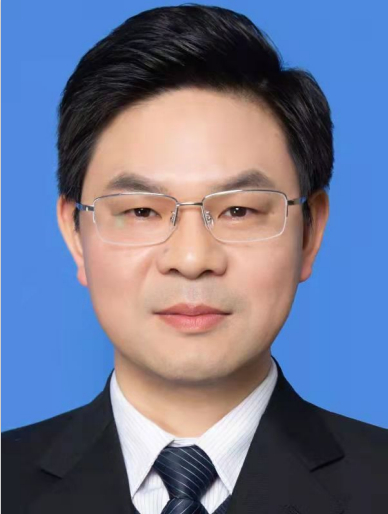}}]{Guang Cheng}
received the B.S. degree in Traffic Engineering from Southeast University in 1994, the M.S. degree in Computer Application from Hefei University of Technology in 2000, and the Ph.D. degree in Computer Network from Southeast University in 2003. He is a Full Professor in the School of Cyber Science and Engineering, Southeast University, Nanjing, China. He has authored or coauthored seven monographs and more than 100 technical papers, including top journals and top conferences. His research interests include network security, network measurement, and traffic behavior analysis. He is a Member of IEEE and a Senior Member of CCF. 
\end{IEEEbiography}

\vfill




\end{document}